%% file: 00_main.tex
\documentclass[preprint, superscriptaddress,amsmath, nofootinbib]{revtex4-1}
\usepackage{graphicx}
\usepackage{latexsym}
\usepackage{slashed}
\usepackage{bm}
\usepackage{color}
\usepackage[colorlinks,citecolor=blue,urlcolor=blue,linkcolor=red]{hyperref}
\usepackage{diagbox}
\usepackage{makecell}
\usepackage{hyperref}
\pdfoutput=1
\usepackage{dcolumn}
\usepackage{bm}
\usepackage{multirow}
\usepackage{subcaption}
\usepackage{ulem}
\usepackage{tabularx}

\begin{document}

\title{Unveiling the Invisible: ALPs and Sterile Neutrinos at the LHC and HL-LHC}
\def\slash#1{#1\!\!\!/}

\author{Kingman Cheung}
\email{cheung@phys.nthu.edu.tw}
\affiliation{Department of Physics, National Tsing Hua University, Hsinchu 30013, Taiwan} 
\affiliation{Center for Theory and Computation,
National Tsing Hua University, Hsinchu 30013, Taiwan}
\affiliation{Division of Quantum Phases and Devices, School of Physics, Konkuk University, Seoul 143-701, Republic of Korea}

\author{C.J. Ouseph}
\email{ouseph444@gmail.com}
\affiliation{Institute of Convergence Fundamental Studies, Seoul National University
of Science and Technology, Seoul 01811, Korea} 

\author{Sin Kyu Kang}
\email{skkang@snut.ac.kr}
\affiliation{Institute of Convergence Fundamental Studies, Seoul National University
of Science and Technology, Seoul 01811, Korea} 
\affiliation{School of Natural Science, Seoul National University
of Science and Technology, Seoul 01811, Korea}

\date{\today}

\begin{abstract}
We investigate the potential of using the signature of mono-Higgs plus large missing energies to constrain on two new physics models, namely 
the model of an axion-like particle (ALP) and the model of sterile neutrinos. We focus on the Higgs-ALP interactions starting at 
dimension-six and the Higgs-sterile neutrino interactions starting at dimension-five, via the processes
$pp \to h a a$ for ALP production and $pp \to h N N$ for sterile neutrinos at the LHC and High Luminosity LHC (HL-LHC),
followed by the Higgs decay $h \to b \bar{b}$.  
We establish bounds on the ALP-Higgs coupling $\frac{C_{aH}}{\Lambda^2}$ and sterile neutrino-Higgs coupling $\frac{\lambda_3}{M_*}$,
respectively, 
for ALP and sterile-neutrino mass ranging from 1 to 60 GeV, using the recent ATLAS data on mono-Higgs plus missing energies
at the LHC $(\sqrt{s} = 13\;{\rm  TeV}\; {\rm and}\;  \mathcal{L} = 139\; {\rm fb}^{-1})$.
The most stringent constraint occurs in the missing transverse energy $M_{ET}$ range $200 < M_{ET} \leq 350$ GeV. 
We also estimate the sensitivities that we can achieve at the HL-LHC ($\sqrt{s} = 14$ TeV and $\mathcal{L} = 3000$ fb$^{-1}$).
We obtain improved sensitivities across various missing energy regions. The ALP model exhibits better sensitivities, 
particularly at lower mass range, compared to the sterile neutrino model, which shows weaker sensitivities across similar mass 
and energy ranges. Our results underscore the potential of the mono-Higgs signature as a robust probe for physics beyond the Standard Model.
\end{abstract}

\maketitle
\input{subtex/01_intro.tex}

\input{subtex/02_model}

\input{subtex/03_exp_simu}

\input{subtex/04_result}

\input{subtex/05_Conclusion}

\section*{Acknowledgment}
 K.C. is supported by the National Science \& Technology Council under grant no. NSTC 113-2112-M-007-041-MY3. C. J. O. and S. K. K are supported by the National Research Foundation of Korea under grant NRF-2023R1A2C100609111.
\begin{appendix}

\input{subtex/Appendix_ALP}

\input{subtex/Appendix_St}

\end{appendix}

\bibliography{paper}

\end{document}

%% file: subtex/01_intro.tex
\section{Introduction}\label{sec.1}

The scalar boson, which was discovered in 2012 \cite{ATLAS:2012yve,CMS:2012qbp}, may hold the key to physics 
Beyond the Standard Model (BSM). It is best described as the Standard Model (SM) Higgs boson, as various couplings of this scalar boson have
been measured with high accuracy (see e.g. \cite{ATLAS:2024fkg,Heo:2024cif}). Yet, quite a few avenues in Higgs
physics may be able to reveal signs of new physics, such as rare decays of the Higgs boson, the self couplings,
strong scattering of the longitudinal component of the massive gauge bosons, or the Higgs boson as the bridge
to hidden or dark sectors. The latter possibility has been intensively carried in the past decade under the 
name Higgs-portal models \cite{Patt:2006fw}, especially associated with 
the dark matter (DM) \cite{Djouadi:2011aa,Arcadi:2019lka}. If the DM particle is lighter than 
half the mass of the Higgs boson, then the Higgs boson can directly decay into a pair of DM particles; 
otherwise the pair of DM particles can be produced via the Higgs propagator. Various types of 
scenarios have been studied in literature.

In this work, we focus on new physics models: (i) an axion-like particle (ALP) $a$ with mass from a few GeV to $m_H/2$,
which couples to the Higgs via a dimension-six operator $(\partial_\mu a)(\partial^\mu a) \phi^\dagger \phi$,
and (ii) a sterile neutrino $N$, which couples to the Higgs via $\phi^\dagger \phi N N$.
We assume that the ALP and the sterile neutrino are stable within the CMS/ATLAS detector such that they will give
rise to missing energy signals if they are produced. This assumption for the ALP remains valid as long as the ALP
couplings to fermions and gauge bosons are sufficiently small so that it can only decay outside the detector.
On the other hand, the sterile neutrino can decay into an active neutrino and a virtual $W$ via its mixing with
the active neutrinos, but the mixing is mostly very small such that it decays outside the detector.
Thus, the final state that we consider in this work is mono-Higgs with large missing energies, followed
by the $H \to b \bar b$ decay. 

A recent result by ATLAS \cite{ATLAS:2021shl} searched for the mono-Higgs plus large missing energies in
various missing energy ranges at the center of energy $\sqrt{s}=13$ TeV and gave limits on the production cross sections $\sigma (pp \to H \bar \chi \chi \to  (b\bar b)\bar \chi \chi)$, where $\chi$ is a Dirac fermion in the hidden sector.
We follow closely their experimental selection cuts and make use of their results 
to put limits on the effective couplings of $(\partial_\mu a)(\partial^\mu a) \phi^\dagger \phi$ and
$\phi^\dagger \phi N N$. We also estimate the sensitivites that one can achieve at the High-Luminosity LHC
with $\sqrt{s}= 14$ TeV and integrated luminosity ${\cal L} = 3000\;{\rm fb}^{-1}$.

The organization is as follows. In the next section, we describe the ALP model and the sterile neutrino model, 
and calculate the production cross sections of the ALP and sterile neutrino, separately.
In Sec.~\ref{sec.3}, we describe the procedures in the simulations and how we follow the selection cuts 
of the ATLAS analysis. We give the results of the limits on the couplings and of the sensitivities 
in Sec.~\ref{sec.4}. Finally, we conclude in Sec.~\ref{sec.5}.

%% file: subtex/02_model.tex
\section{Highlights of the Models}

\subsection{Axion-like Particke}\label{sec.2a} 

The effective interactions of the ALP $a$ with the SM fields emerge at dimension-5~\cite{Georgi:1986df,Bauer:2018uxu}:
\begin{eqnarray}
  {\cal L}^{D=5} &=& \frac{1}{2} (\partial_\mu a) (\partial^\mu a) 
  - \frac{1}{2} m_a^2 a^2
  + \sum_f \frac{C_{ff}}{2\Lambda} \partial^\mu a \,
  \bar f \gamma_\mu \gamma_5 f  \nonumber \\
  &+& g_S^2 \frac{C_{GG}}{\Lambda} a G^A_{\mu\nu} \tilde{G}^{\mu\nu,A}
  + g^2  \frac{C_{WW}}{\Lambda} a W^i_{\mu\nu} \tilde{W}^{\mu\nu,i}
  +g'^2 \frac{C_{BB}}{\Lambda} a B_{\mu\nu} \tilde{B}^{\mu\nu} \;,
\end{eqnarray}
where $A = 1,\ldots,8$ represents the $SU(3)$ color indices, and $i = 1,2,3$ labels the $SU(2)$ indices. The gauge couplings $g_S$, $g$, and $g'$ correspond to the groups $SU(3)$, $SU(2)$, and $U(1)_Y$, respectively.
In this framework, we set $C_{GG} = 0$ to prevent ALP-QCD axion mixing, thereby avoiding the reappearance of the strong CP problem. Following the rotation of $B_\mu$ and $W^3_\mu$ into the physical photon and $Z$ boson, the ALP interactions with the photon and $Z$ boson are given by:
\begin{eqnarray}
   {\cal L} &=& e^2 \frac{C_{\gamma\gamma}}{\Lambda} a F_{\mu\nu} \tilde{F}^{\mu\nu}
   + \frac{2e^2}{s_w c_w} \frac{C_{\gamma Z}}{\Lambda} a F_{\mu\nu} \tilde{Z}^{\mu\nu}
   + \frac{e^2}{s_w^2 c_w^2} \frac{C_{ZZ}}{\Lambda} a Z_{\mu\nu} \tilde{Z}^{\mu\nu} \;, \label{eq:gamma}
\end{eqnarray}
where the coefficients are defined as:
\[
  C_{\gamma\gamma} = C_{WW} + C_{BB}\;, \;\;\;
  C_{\gamma Z } = c_w^2 C_{WW} - s_w^2 C_{BB}\;, \;\;\;
  C_{ZZ} = c_w^4 C_{WW} + s_w^4 C_{BB}\;,
\]
and $s_w$ and $c_w$ denote the sine and cosine of the weak mixing angle, respectively.

The interactions between the ALP and the Higgs boson arise at dimension-6:
\begin{equation}
  \label{lag}
   {\cal L}^{D \geq 6}
   = \frac{C_{aH}}{\Lambda^2} (\partial_\mu a) (\partial^\mu a) \,
   \phi^\dagger \phi
   + \frac{C_{aZH}}{\Lambda^3} (\partial^\mu a) \,
   \left( \phi^\dagger i D_\mu \phi + {\rm h.c.} \right )\,
   \phi^\dagger \phi \;,
 \end{equation}
where the covariant derivative $D_\mu$ is given by:
\[
 D_\mu = \partial_\mu + i \frac{g}{\sqrt{2}}
 \left( W^+_\mu \tau^+ + W^-_\mu \tau^- \right ) + i e Q A_\mu
 + i\frac{g}{c_w} (T_3 - s_w^2 Q ) Z_\mu \;,
\]
with $\tau^\pm$ being the $SU(2)$ ladder operators, $T_3$ the third component of weak isospin, and $Q$ the electric charge.

\par In this study, we consider ALP as a stable particle candidate that does not decay into any other particles. In this context, we set all ALP couplings to zero, specifically $C_{ff}$, $C_{GG}$, $C_{WW}$, $C_{BB}$, and $C_{aZH}$, except for the ALP-Higgs coupling $C_{aH}$. The relevant effective Lagrangian for this analysis can be written as follows:

\begin{equation}
  \label{lag_f}
   {\cal L}^{D \geq 6}
   = \frac{C_{aH}}{\Lambda^2} (\partial_\mu a) (\partial^\mu a) \,
   \phi^\dagger \phi.
 \end{equation}

\subsubsection{Production of Invisible Axion-Like Particles with a Mono-Higgs Signature}\label{sec.2a1}

\begin{figure}
    \centering
    \begin{subfigure}[b]{0.48\textwidth}
        \centering
        \includegraphics[width=\textwidth]{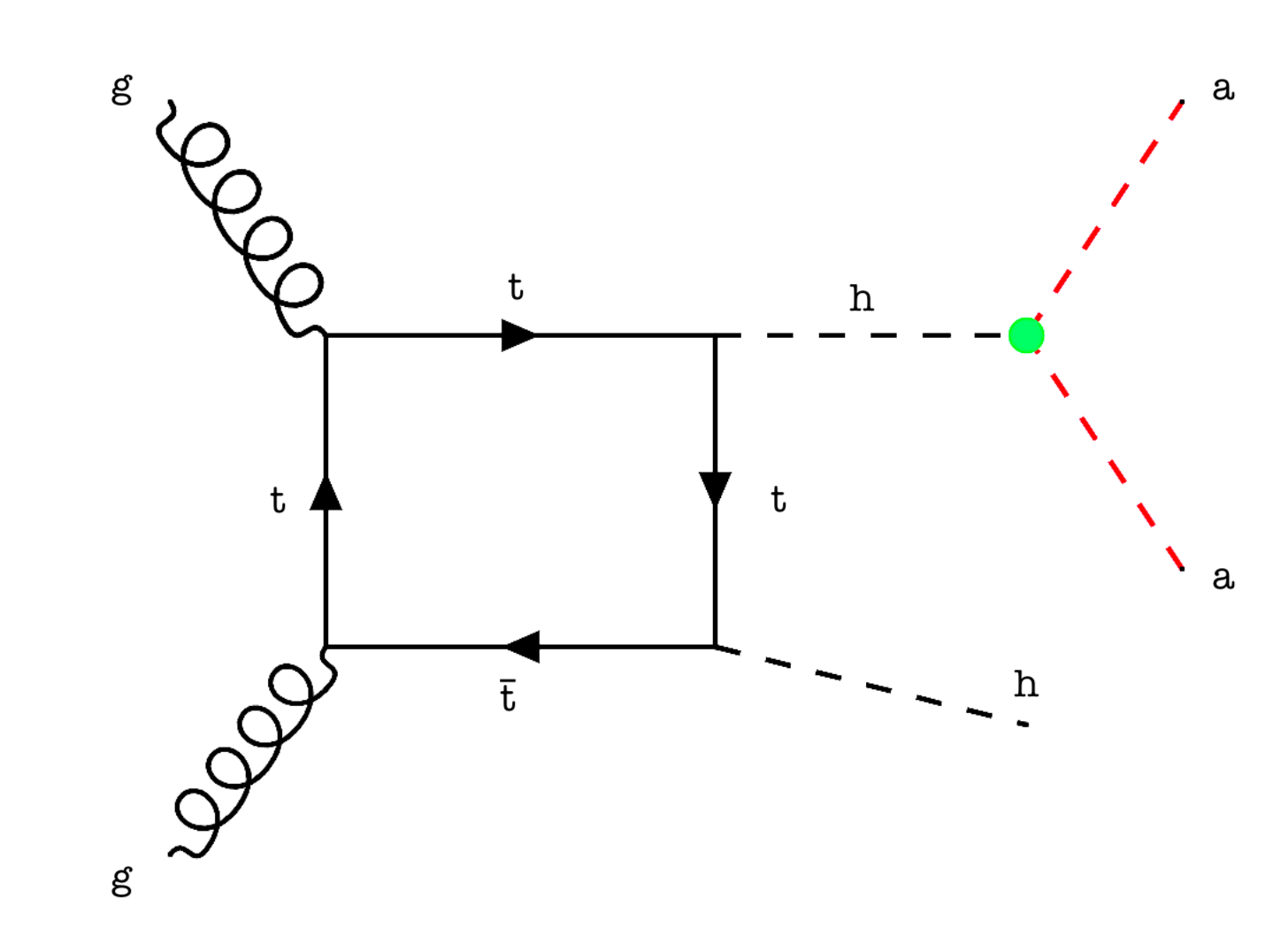}
        \caption{}
        \label{fig.alp_1}
    \end{subfigure}
    \hfill
    \begin{subfigure}[b]{0.51\textwidth}
        \centering
        \includegraphics[width=\textwidth]{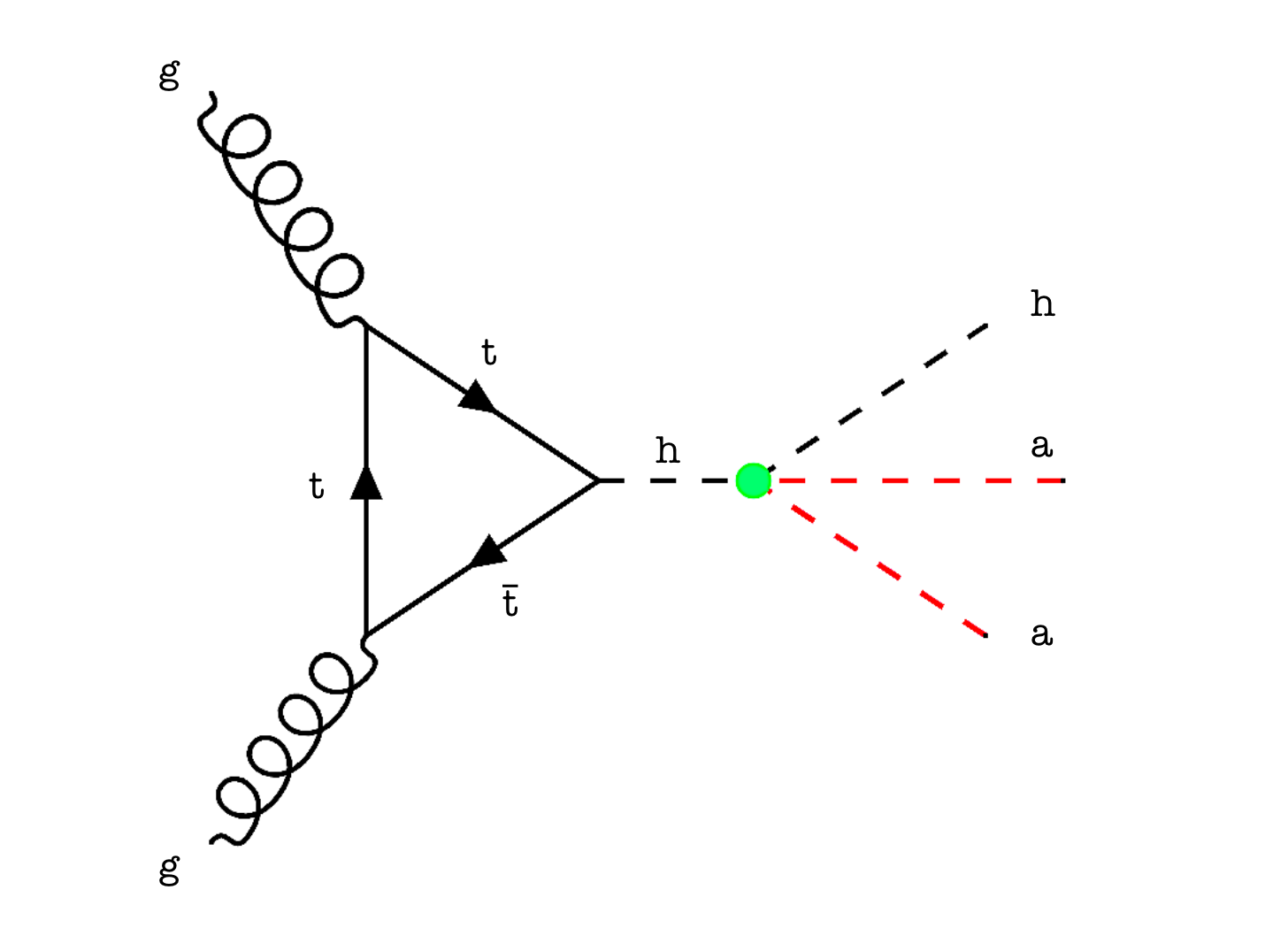}
        \caption{}
        \label{fig.alp_2}
    \end{subfigure}
    \hfill
    \begin{subfigure}[b]{0.45\textwidth}
        \centering
        \includegraphics[width=\textwidth]{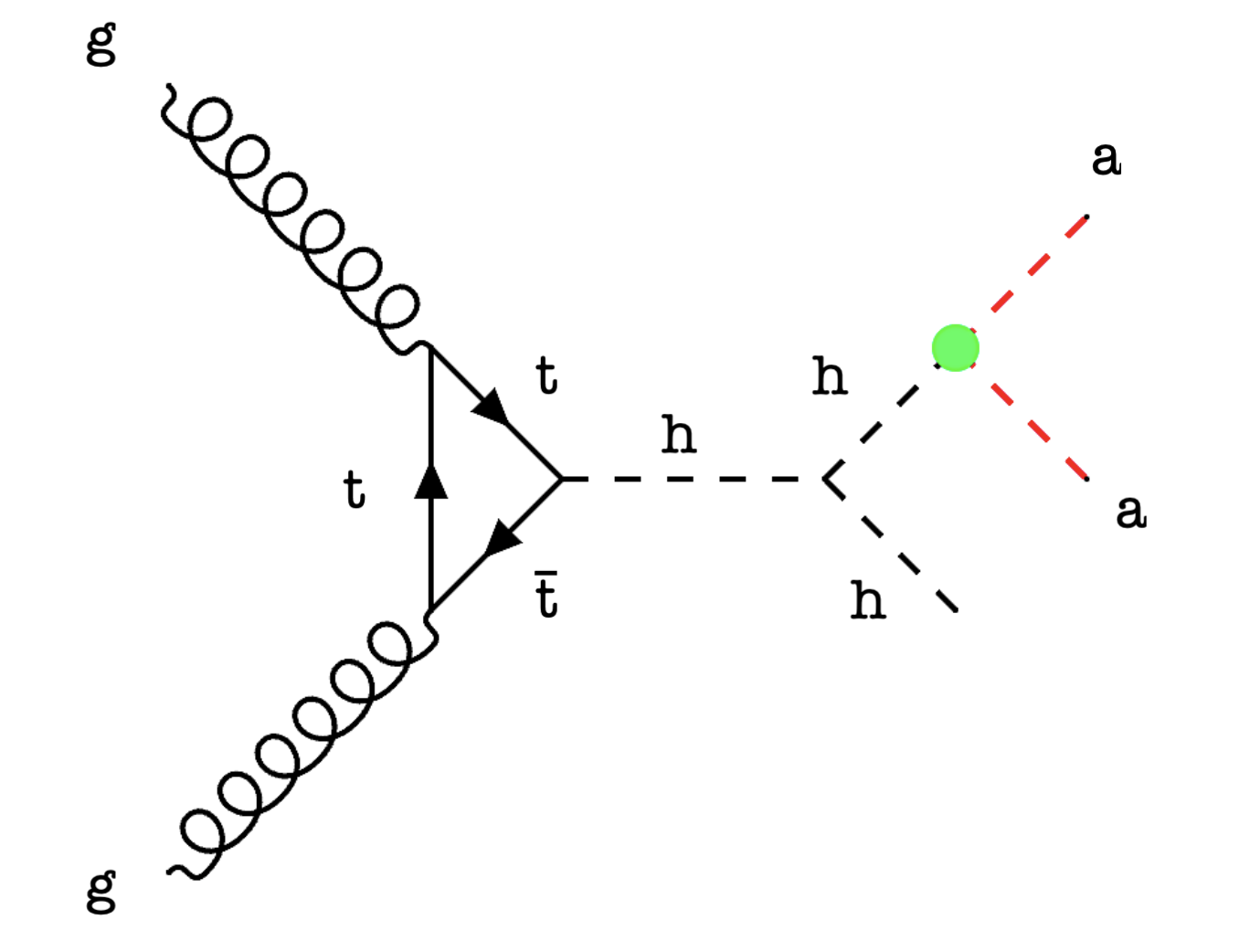}
        \caption{}
        \label{fig.alp_3}
    \end{subfigure}
    \caption{Relevant Feynman diagrams for the process $pp \to h a a$, where $h$ is the Higgs boson and $a$ represents an ALP.}
    \label{fig.alp_feyn}
\end{figure}

Our focus is to explore the \(\frac{C_{aH}}{\Lambda^2}\) coupling by studying the mono-Higgs signature produced alongside the ALPs 
in the process \(pp \to h a a\) at the LHC with \(\sqrt{s} = 13\) TeV and at the HL-LHC with \(\sqrt{s} = 14\) TeV. 
The relevant Feynman diagrams for the process \(pp \to h a a\) are shown in FIG.~\ref{fig.alp_feyn}. 
We used \texttt{MadGraph5\_aMC@NLO}~\cite{Alwall:2011uj} to calculate the production cross sections 
for the process \(pp \to h a a\), fixing the parameters \(C_{aH} = 1\) and \(\Lambda = 1000\) GeV. 
For the \texttt{MadGraph5\_aMC@NLO} simulation process, we built a UFO file for the Lagrangian quoted 
in Eq.~(\ref{lag_f}) by modifying the general \texttt{ALP\_UFO}~\cite{Brivio:2017ije}\footnote{This UFO model file is publicly accessible for download at \url{https://feynrules.irmp.ucl.ac.be/wiki/ALPsEFT}} file using the Mathematica package \texttt{FeynRules}~\cite{Alloul:2013bka}.

\begin{figure}
    \centering
    \includegraphics[width=0.7\textwidth]{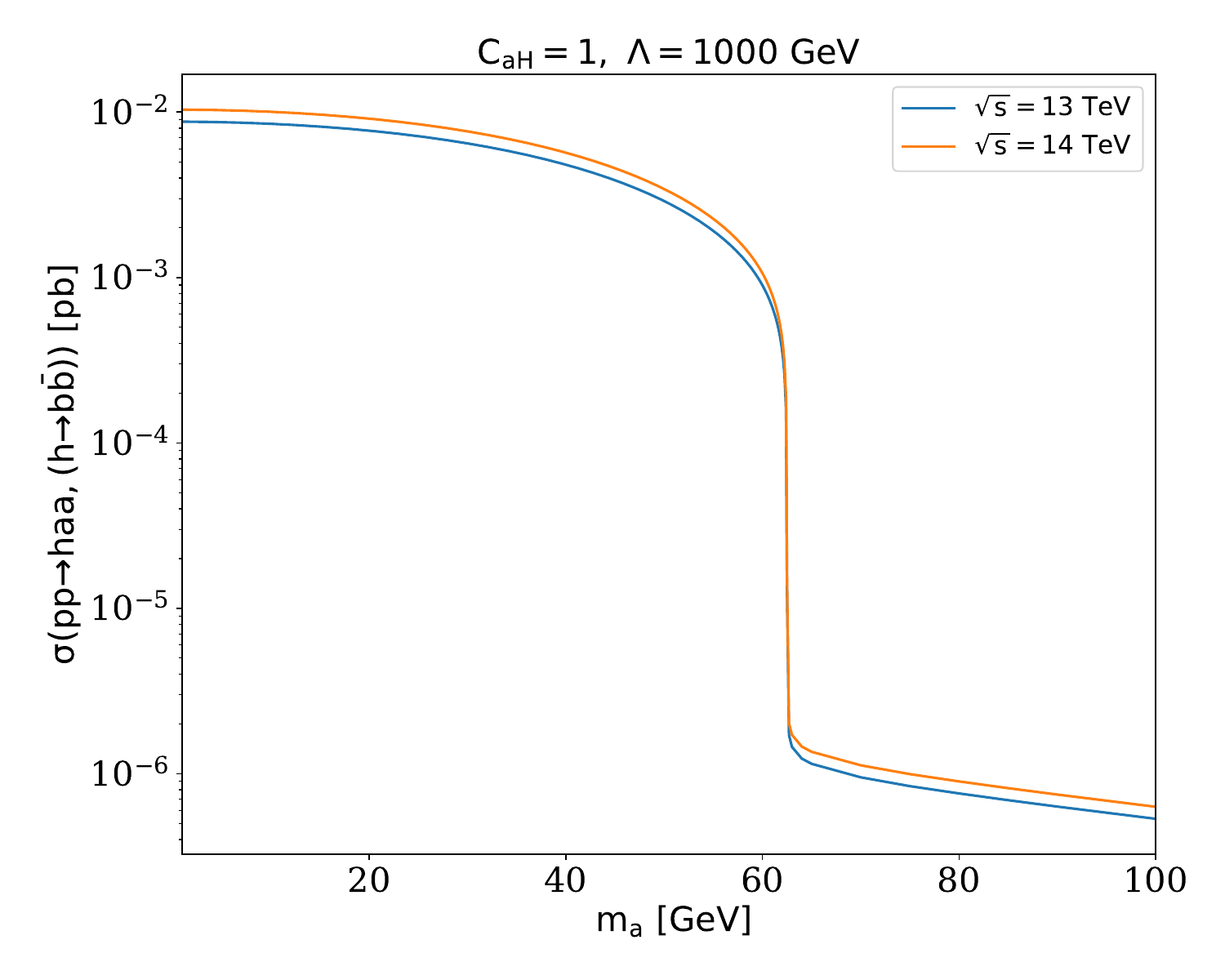}
    \caption{ALP production cross section $\sigma(pp \to h a a, (h \to b\bar{b}))$ at the LHC with $\sqrt{s}=13$ and 14 TeV. The coupling is fixed by
    setting $C_{aH}=1$ and $\Lambda = 1000$ GeV.
    The uncertainty associated 
with the choice of renormalization scale is estimated to be about $20-25\%$.} 
    \label{fig.cross}
\end{figure}

The total cross section for the signal process \(pp \to h a a\) and the subsequent decay of the Higgs to a pair of \(b\) quarks \(h \to b \bar{b}\) is evaluated using the formula given by
\begin{equation}\label{Eq.cross}
    \sigma(pp\to h a a, (h\to b\bar{b}))=\sigma (pp\to h a a)\times \mathcal{B}(h\to b\bar{b}).
\end{equation}
The branching ratio of the Higgs to a pair of \(b\) quarks is \(\mathcal{B}(h \to b \bar{b}) = 0.582\)\footnote{It is available at \texttt{\url{https://twiki.cern.ch/twiki/bin/view/LHCPhysics/CERNYellowReportPageBR}}.}. 
The estimated cross section for the signal process at both the LHC (\(\sqrt{s} = 13\) TeV) and the HL-LHC (\(\sqrt{s} = 14\) TeV) for an ALP mass ranging from 
\(1~\text{GeV} \leq m_a \leq 100~\text{GeV}\) 
\footnote{The lower end of the mass range is, in principal, can be further lowered to a very 
small value, say keV, but there are strong cosmological and astrophysical limits on ALP for
$m_a \alt 1$ GeV. See for example
\url{[https://cajohare.github.io/AxionLimits/docs/fips.html]} Therefore, we stay away from such constraints in our study.}
is depicted in FIG.~\ref{fig.cross}. This kind of loop-induced process can be initiated in \texttt{MadGraph5\_aMC@NLO} by switching on \texttt{[QCD]}~\cite{Hirschi:2015iia}, and the subsequent decay of \(h \to b \bar{b}\) can be initiated using \texttt{MadSpin}~\cite{Artoisenet:2012st}.

From Fig.~\ref{fig.cross} we can see a sharp drop in cross section around 60 GeV. It is
obvious that when $m_a < m_H/2$ both Higgs bosons are produced on-shell; but when 
$m_a \ge m_H/2$ one of the Higgs boson is forced to be off-shell, which results in
the sharp drop of production. We therefore focus on the mass range
$1\,{\rm GeV} < m_a < 60 \,{\rm GeV}$ in our analysis.

\begin{figure}
     \centering
    \includegraphics[width=0.7\textwidth]{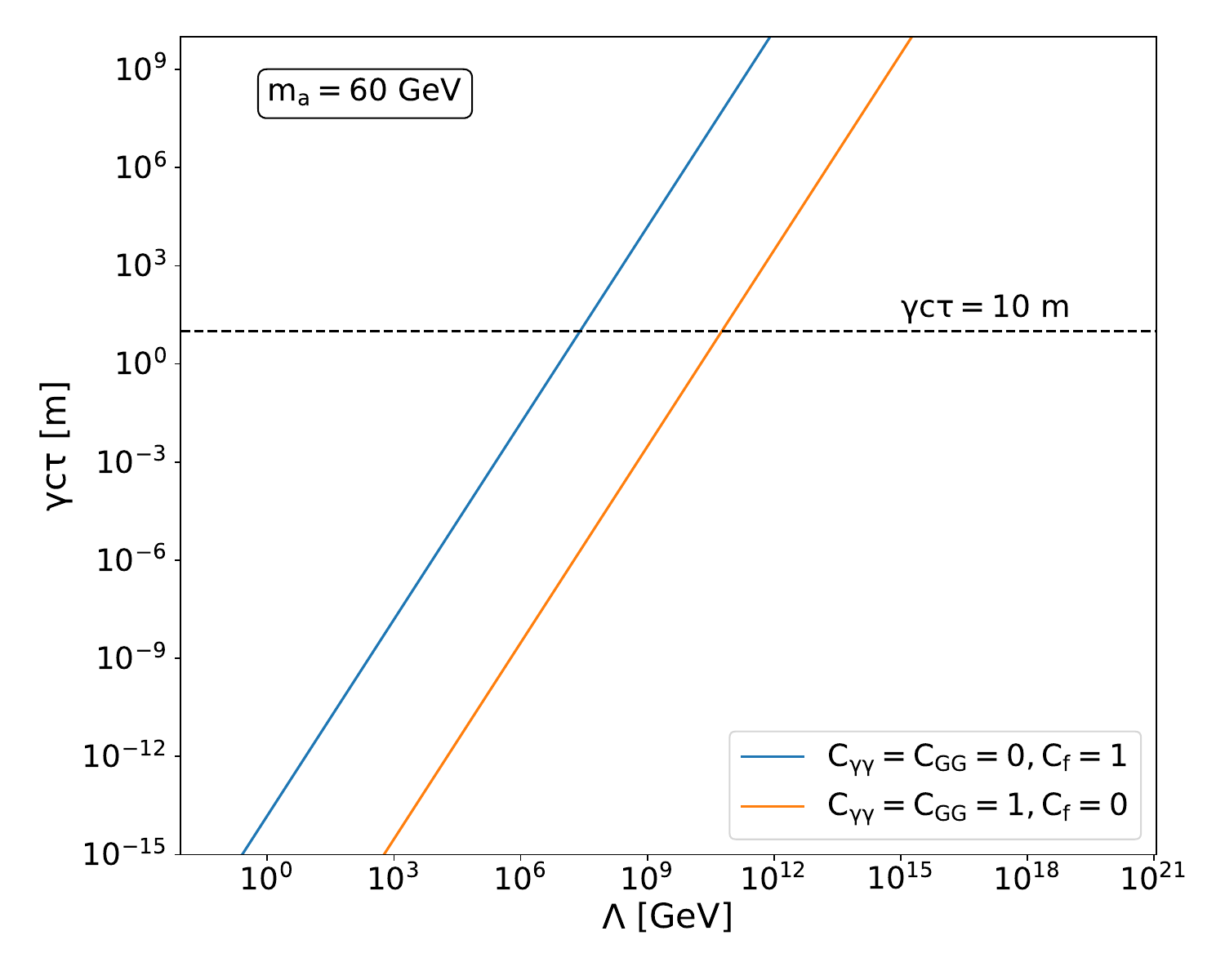}
    \caption{Decay length $\gamma c\tau $ of the ALP versus the scale $\Lambda$. A typical size $O(10)$ m (horizontal line) is shown. ALP Partial Decay widths are available in \cite{Bauer:2018uxu}}
    \label{decay}
\end{figure}

\subsubsection{Decay length of the Axion-like Particle}
In this analysis, we simply assume that the ALP is stable on the collider scale, i.e., its 
decay length is longer than the typical size $O(10)$ m of a detector. In reality, the ALP may 
afford tiny values of the ALP-gauge and ALP-fermion couplings such that it still decays outside 
the detector.  We calculate the decay length $\gamma c \tau $ of the ALP, where $\gamma = E_a /m_a$
and $E_a \simeq m_H/2$. Here $\tau \equiv 1/\Gamma_a$ is the inverse of the total decay width
of the ALP. The total decay width of the ALP is calculated by considering two scenarios:  

\begin{enumerate}
    \item In the first scenario, we sum the partial widths of the decay channels $a \to \gamma \gamma$ and $a \to gg$, assuming that the ALP couplings to electrons and muons are zero. We set the ALP-gauge boson couplings to unity, i.e., $C_{\gamma\gamma} = C_{GG} = 1$ and $C_f = 0$. The corresponding decay length is represented by the orange line in Fig.~\ref{decay}. For this case, we find that $\Lambda \sim 10^{11}$ GeV for $\gamma c \tau \sim O(10)$ meters.  

    \item In the second scenario, we sum the partial widths of the decay channels $a \to e^+e^-$ and $a \to \mu^+\mu^-$, assuming that the ALP couplings to photons and gluons are zero while setting the ALP-fermion coupling to unity, i.e., $C_{\gamma\gamma} = C_{GG} = 0$ and $C_f = 1$. For this case, we find that $\Lambda \sim 10^{7}$ GeV for $\gamma c \tau \sim O(10)$ meters.  
\end{enumerate}  

For lighter ALPs, the decay length increases as the mass decreases. Therefore, to ensure that the ALP remains intact within the detector, a typical upper bound on the coupling is $C/\Lambda \simeq 10^{-11}\, {\rm GeV}^{-1}$.

\subsection{Sterile Neutrino}\label{sec.6}
The Lagrangian presented in Eq.~(\ref{Eq.sterile_L}) describes the interactions involving a sterile neutrino \( N \), with each term corresponding to a distinct type of coupling \cite{Graesser:2007yj,Graesser:2007pc}:
\begin{equation}\label{Eq.sterile_L}
    \mathcal{L} = \lambda_1 M_* N N + \lambda_2 L H N + \frac{\lambda_3}{M_*} H^{\dag} H N N \;.
\end{equation}
The first term, \( \lambda_1 M_* N N \), introduces a Majorana mass for the sterile neutrino \( N \). Here, \( \lambda_1 \) is a dimensionless coupling constant and \( M_* \) denotes a cut-off scale. This term generates a mass for \( N \) and violates lepton number by two units, a characteristic feature of Majorana masses.
The second term, \( \lambda_2 L H N \), represents the mixing  between active and sterile neutrinos. 
In this term, \( L \) is the lepton doublet, \( H \) is the Higgs doublet, and \( \lambda_2 \) is another dimensionless coupling constant. This coupling allows for possible oscillations between active and sterile neutrino states.
The third term, \( \frac{\lambda_3}{M_*} H^{\dag} H N N \), describes the interaction between 
the Higgs field \( H \) and the sterile neutrino \( N \), and \( \lambda_3 \) is a dimensionless coupling constant.
This higher-dimensional operator is suppressed by the scale \( M_* \) and may be relevant within certain 
effective field theories.
Each of these terms provides insights into sterile neutrino physics, contributing to mass generation, mixing with active neutrinos, and potential interactions with the Higgs field. 
After electroweak symmetry breaking, the first and third terms of the Lagrangian (\ref{Eq.sterile_L}) generate the sterile neutrino mass term, whereas the second term induces a mixing mass between active and sterile neutrinos.
As a result of this mixing mass term, the flavor eigenstates of the neutrino sector do not align with their mass eigenstates, as expressed below;
\begin{equation}
   \nu_{\alpha}=\sum_{i=1}^3U_{\alpha i}\nu_i+U_{\alpha N}N^m, 
\end{equation}
where the index $\alpha$ denote the flavor $e,\mu,\tau$ and $N$, and $N^m$ represents the mass eigenstate of the sterile neutrino.
In the limit $U_{\alpha N} \rightarrow 0$, the mass eigenstate $N^m$ becomes identical to the flavor state $N$.  In this study, we focus on the regime where \( \frac{\lambda_3}{M_*} \gg \lambda_2 \), making the effects of the higher-dimensional operator dominant over the mixing term.


In this study, we consider sterile neutrinos as long lived particles that do not decay into other particles within the LHC detector. This assumption is valid as long as $\lambda_2$ is below $10^{-7}$, above which sterile neutrinos may decay into three active neutrinos via 3-body decay processes mediated by $Z^0$ inside the detector.  Such a small $\lambda_2$ results in a very small mixing between active and sterile neutrinos, and thus $N^m$ can be identified with $N$. Additionally, the decay of Higgs into a pair of light neutrinos is negligibly suppressed.
The relevant effective Lagrangian for this analysis is then simplified to:

\begin{equation}\label{Eq.sterile_L1}
    \mathcal{L} = 
\frac{\lambda_3}{M_*} H^{\dag} H N N
\end{equation}

\subsubsection{Production of Invisible Sterile Neutrino with a Mono-Higgs Signature}\label{sec.2b1}

\begin{figure}[h!]
    \centering
    \includegraphics[width=0.7\textwidth]{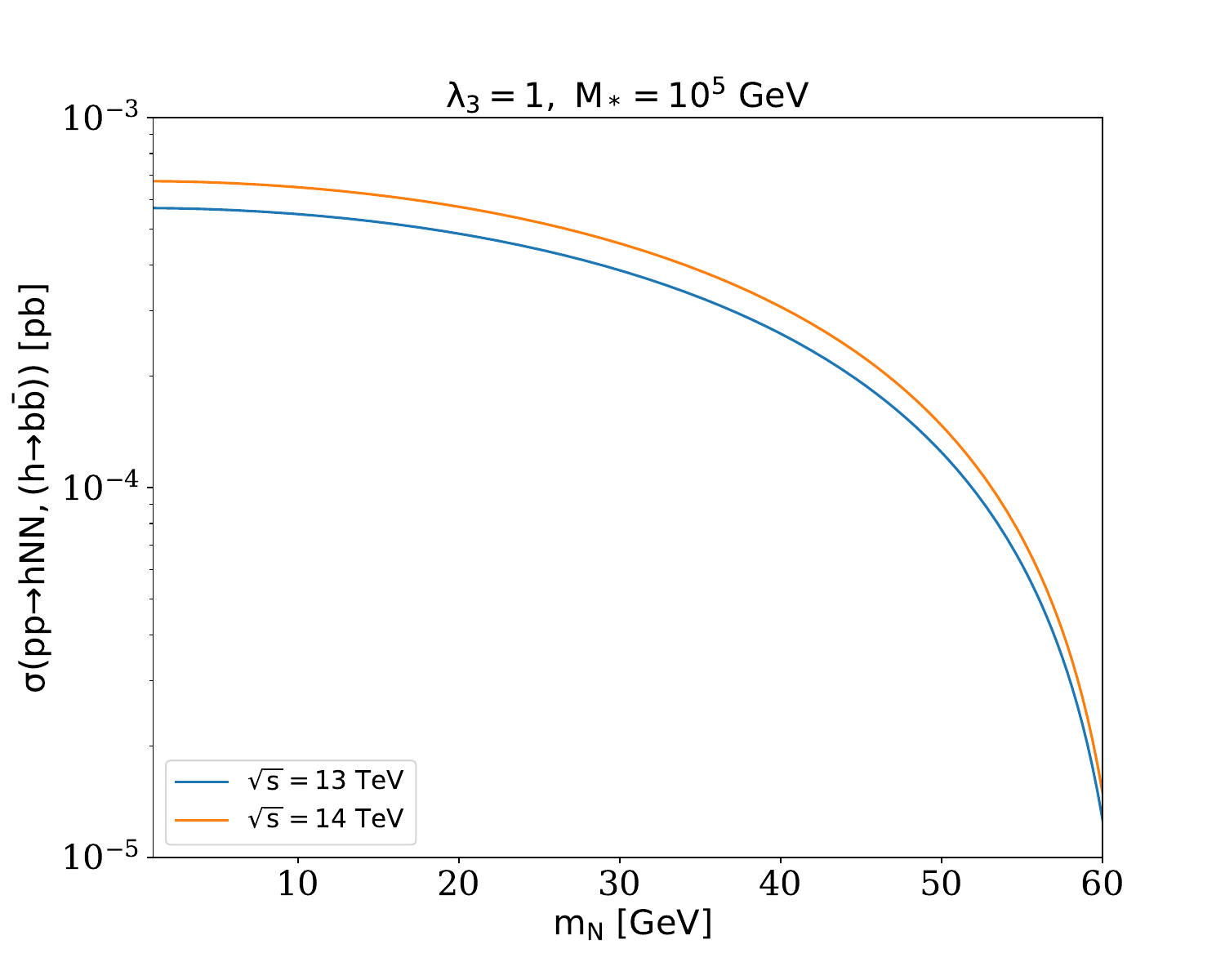}
    \caption{Sterile neutrino production cross section $\sigma(pp \to h N N, (h \to b\bar{b}))$ at the LHC with $\sqrt{s}=13$ and 14 TeV. 
    The coupling is fixed by setting $\lambda_3=1$ and $M_* = 10^5$ GeV. The uncertainty associated 
with the choice of renormalization scale is estimated to be about $20-25\%$.}
    \label{fig.cross_N}
\end{figure}

Our objective is to study the \(\frac{\lambda_3}{M_*^2}\) coupling by analyzing the mono-Higgs signature generated in association with invisible sterile neutrinos in the process \(pp \to h N N\) at the LHC with \(\sqrt{s} = 13\) TeV and at the HL-LHC with \(\sqrt{s} = 14\) TeV. 
The Feynman diagrams relevant to this process \(pp \to h N N\) shares the same topology as the previous 
ALP process and thus similar to those in FIG~\ref{fig.alp_feyn}. 
To calculate the production cross sections for the process \(pp \to h N N\), we employed \texttt{MadGraph5\_aMC@NLO}~\cite{Alwall:2011uj}, 
setting the parameters \(\lambda_3 = 1\) and \(M_* = 10^5\) GeV. 
For this simulation in \texttt{MadGraph5\_aMC@NLO}, we created a UFO file corresponding to 
the Lagrangian defined in Eq.~(\ref{lag_f}) by modifying the generic \texttt{Sterile\_Neutrino\_UFO}~\cite{Degrande:2016aje,Pascoli:2018heg,Cirigliano:2021peb,Alva:2014gxa,Atre:2009rg}\footnote{The UFO model file can be downloaded from \url{https://feynrules.irmp.ucl.ac.be/wiki/HeavyN}} using the Mathematica package \texttt{FeynRules}~\cite{Alloul:2013bka}.

The total cross section for the signal process \(pp \to h N N\), followed by the subsequent Higgs decay to a \(b\)-quark pair \(h \to b \bar{b}\) is calculated using the formula given by
\begin{equation}\label{Eq.cross}
    \sigma(pp\to h N N, (h\to b\bar{b})) = \sigma (pp\to h N N) \times \mathcal{B}(h \to b \bar{b}).
\end{equation}
 The branching ratio of the Higgs decaying to \(b\)-quarks is set as \(\mathcal{B}(h \to b \bar{b}) = 0.582\)\footnote{Available at \texttt{\url{https://twiki.cern.ch/twiki/bin/view/LHCPhysics/CERNYellowReportPageBR}}.}. FIG~\ref{fig.cross_N} 
 displays the cross section for the signal process at both the LHC (\(\sqrt{s} = 13\) TeV) and the HL-LHC (\(\sqrt{s} = 14\) TeV) 
 across the sterile neutrino mass from \(1~\text{GeV} \leq m_N \leq 60~\text{GeV}\). 
 This loop-induced process can be initiated in \texttt{MadGraph5\_aMC@NLO} by 
 enabling \texttt{[QCD]}~\cite{Hirschi:2015iia}, and the subsequent Higgs decay \(h \to b \bar{b}\) is managed using \texttt{MadSpin}~\cite{Artoisenet:2012st}.

%% file: subtex/03_exp_simu.tex
\section{Experiment and Simulations}\label{sec.3}

\subsection{ALP Signal and Associated SM Background Processes} \label{sec.3a}

In this section, we outline the methodology and experimental framework used to differentiate the signal from the primary SM backgrounds. 
We present the event rates for $\sqrt{s} = 13$~TeV with an integrated luminosity of $139~\text{fb}^{-1}$ in order to apply the results 
from ATLAS \cite{ATLAS:2021shl}, and for $\sqrt{s} = 14$~TeV with $3000~\text{fb}^{-1}$ (High-Luminosity LHC~\cite{Apollinari:2015bam}) 
to estimate the sensitivities. 

For the first part we closely follow the experimental cuts outlined in the ATLAS paper~\cite{ATLAS:2021shl} 
to directly make use of their model-independent upper limits on the signal cross-sections in the various energy-missing regions. 
The Monte Carlo simulations for the signal events are performed using \texttt{MadGraph5\_aMC@NLO}. The UFO model 
file corresponding to the ALP Effective Field Theory (EFT) framework (Eq.~(\ref{lag_f})) is utilized for 
simulating signal events. In this simulation, \(5 \times 10^4\) events are generated for the signal process.
We include parton showering and hadronization with \texttt{Pythia8}~\cite{Sjostrand:2007gs} and 
detector simulations using \texttt{Delphes3}~\cite{deFavereau:2013fsa}. For precision and consistency
the \texttt{ATLAS\_card.dat} is incorporated. A jet-cone size of \(R = 0.4\) is used for jet clustering via 
the FastJet package~\cite{Cacciari:2011ma} with the \(anti\)-\(k_T\) algorithm~\cite{Cacciari:2008gp}. 
The output root files from \texttt{Delphes3} are then analyzed further using the 
Python-based tool \texttt{uproot}~\cite{uproot5}.

The final state consists of a pair of invisble ALPs mainly from one of the Higgs bosons and a pair of
$b$ quarks from the decay of the other Higgs boson. Thus, we focus on the ALP with the 
mass range from \(1~\text{GeV}\) to \(60~\text{GeV}\). We choose typical input values for 
the model parameters: \(\Lambda = 1000~\text{GeV}\) and \(C_{aH} = 1\), such that the event rate is scaled as 
$(C_{aH}/\Lambda^2)^2$.
The number of signal events $N_s$ is calculated as
\begin{equation}\label{eq:Event_rate}
        N_s = \sigma_{s} \times \frac{N_{\text{selected}}}{N_{\text{sim}}} \times \mathcal{L}.
\end{equation}
Here $N_{selected}$ denotes the number of events that pass the selection cuts, $N_{sim}$ is the total number of simulated events
and thus $\frac{N_{\text{selected}}}{N_{\text{sim}}}$ denotes the selection efficiency, and $\mathcal{L} $ 
is the integrated luminosity.
The signal cross section $\sigma(pp \to h aa,\; h\to b \bar b)$ is shown in FIG.~\ref{Eq.cross}. 

\subsubsection{Event selection}\label{sec.3a1}
For event selection we adhere to the cuts defined in the resolved region outlined in Ref.~\cite{ATLAS:2021shl}.
This region selects events with \(M_{ET} < 500\) GeV and requires at least two \(b\)-tagged small-\(R\) jets, 
with the two highest \(p_T\) jets forming the Higgs boson candidate. The resolved region retains a reasonable number of 
signal events after selection. 
The ATLAS paper also defined a merged region, which selected events with \(M_{ET} > 500\) GeV. 
In this region, at least one large-\(R\) jet is required, with the two leading variable-\(R\) 
track jets associated with the leading large-\(R\) jet being \(b\)-tagged. 
However, since the Merged Region requires \(M_{ET} > 500\) GeV, this cut significantly reduces the number 
of events in our scenario. The signal event rate in the merged Region for the HL-LHC ($\sqrt{s}=14$ TeV, 
$\mathcal{L}=3000~\mathrm{fb^{-1}}$) is comparatively higher than that of 
the LHC ($\sqrt{s}=13$ TeV, $\mathcal{L}=139~\mathrm{fb^{-1}}$). The detailed results for the merged region 
are presented in Appendix~\ref{App1}. In the main text, we focus on the resolved region.

\begin{table}[t!]
\centering
\begin{ruledtabular}
\begin{tabular}{lccc}
\textbf{Cut} & \textbf{$m_a = 1$ GeV} & \textbf{$m_a = 30$ GeV} & \textbf{$m_a = 60$ GeV} \\
\hline
Total no. of events                               & 1181.40  & 899.45   & 124.86 \\
$M_{ET} > 150\,{\rm GeV}$                            & 538.29   & 406.82   & 56.72  \\
Lepton veto                                & 538.18   & 406.77   & 56.70  \\
Muon veto                                  & 537.82   & 406.41   & 56.65  \\
Tauon veto                                 & 505.24   & 380.34   & 53.02  \\
$\Delta \phi(jet_{123}, M_{ET}) > 20^\circ$ & 463.63   & 348.60   & 48.56  \\
$M_{ET} < 500\,{\rm GeV}$                             & 454.49   & 341.16   & 47.55  \\
$\geq 2$ b-tagged small-R jets             & 172.06   & 130.69   & 17.88  \\
$P_T^h$ selection                          & 146.61   & 111.46   & 15.19  \\
$m_T^{b,\min} > 170$ GeV                   & 58.76    & 43.61    & 6.03   \\
$m_T^{b,\max} > 200$ GeV                   & 36.88    & 27.13    & 3.80   \\
$N_{\text{jets}}$ selection                & 25.73    & 19.61    & 2.63   \\
$m_{b\bar{b}}$ mass selection              & 23.68    & 17.85    & 2.36   \\
\hline
\textbf{2-btag} & & & \\
$150 \leq M_{ET}<200$ GeV                     & 6.62     & 5.36     & 0.69   \\
$200 \leq M_{ET}<350$ GeV                    & 14.81    & 10.88    & 1.48   \\
$350 \leq M_{ET}<500$ GeV                    & 1.51     & 1.22     & 0.13   \\
\hline
\textbf{3-btag} & & & \\
$150 \leq M_{ET}<200$ GeV                    & 0.19     & 0.09     & 0.01   \\
$200 \leq M_{ET}<350$ GeV                    & 0.47     & 0.25     & 0.05   \\
$350 \leq M_{ET}<500$ GeV                    & 0.07     & 0.04     & 0.00   \\
\end{tabular}
\end{ruledtabular}
\caption{Cut-flow table for various values of $m_a$ in the resolved Region $M_{ET}<500$ GeV for the LHC ($\sqrt{s}=13$ TeV, $\mathcal{L}=139~\mathrm{fb^{-1}}$).
The coupling is fixed by setting $C_{aH}=1$ and $\Lambda = 1000$ GeV. The numbers of events are calculated by Eq.~(\ref{eq:Event_rate}).
\footnote{ The b-tagging efficiency employed in Ref.~\cite{ATLAS:2021shl} was based 
on Ref.~\cite{ATLAS:2019bwq}. }
}
\label{cutflow13}
\end{table}

\begin{figure}
    \centering
    \begin{subfigure}[b]{0.49\textwidth}
        \centering
        \includegraphics[width=\textwidth]{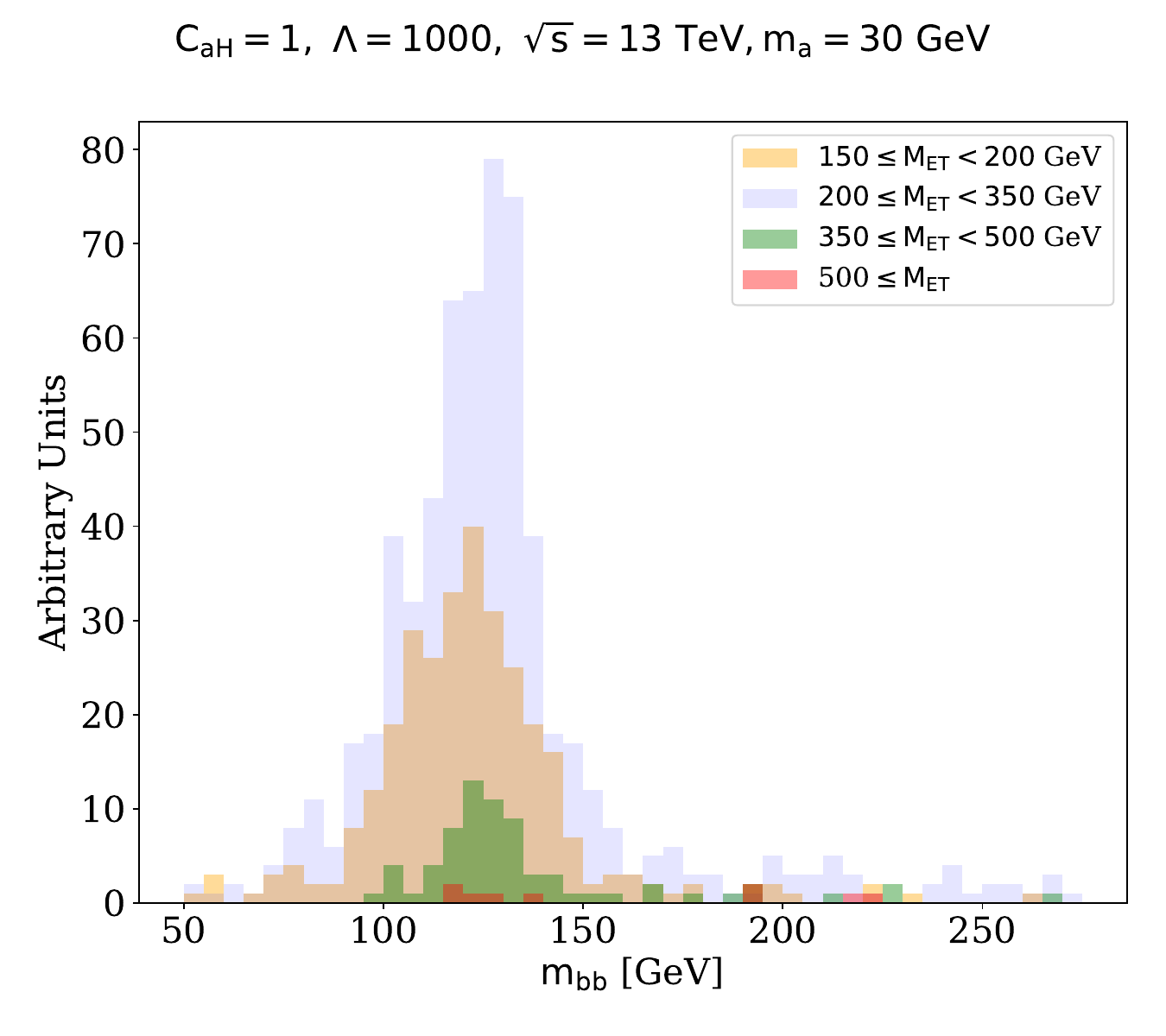}
        \label{fig.dis_13TeV}
    \end{subfigure}
    \hfill
    \begin{subfigure}[b]{0.49\textwidth}
        \centering
        \includegraphics[width=\textwidth]{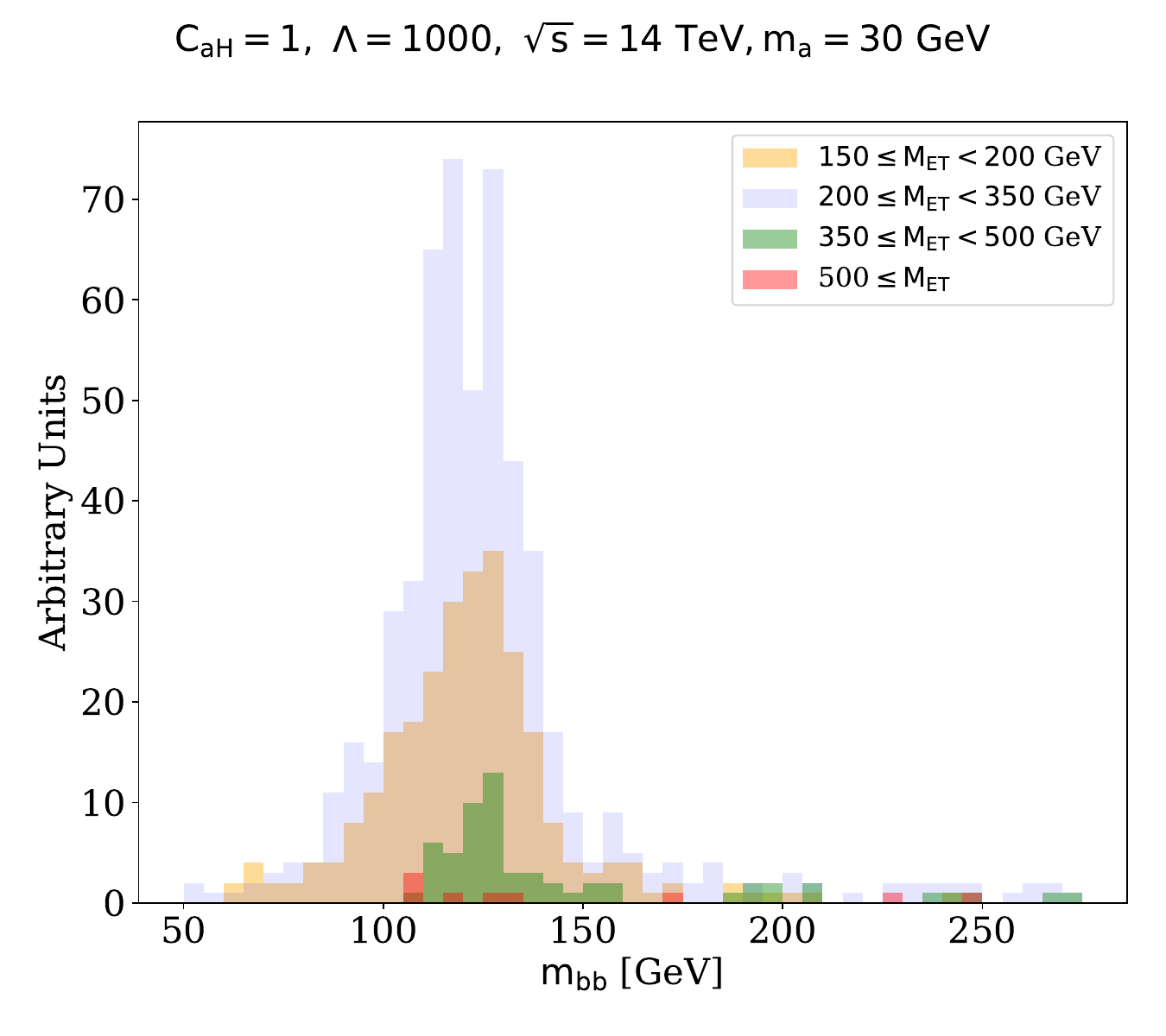}
        \label{fig.dis_14TeV}
    \end{subfigure}
    \caption{Invariant mass $m_{bb}$ distributions for the $b$-quark pair in the process
    $pp \to h a a\,;\, h \to b \bar b$ at the LHC with (a) $\sqrt{s}=13$ TeV and (b) $\sqrt{s}=14$ TeV in various missing energy regimes.}
    \label{fig.mbb}
\end{figure}
\par The Event selection cuts are given below.

\begin{itemize}
    \item \textbf{Missing transverse energy, $M_{ET} > 150$ GeV:} \\
    A large missing transverse energy (\(M_{ET}\)) in the event is a key indicator of undetectable particles, such as dark matter candidates, escaping the detector. Events with \(M_{ET} > 150\) GeV are selected to target processes where invisible particles are produced, as high \(M_{ET}\) is typically associated with missing momentum carried away by these particles.

    \item \textbf{Lepton veto:} \\
    To reduce background contamination from the SM processes involving leptonic decays, such as the decays of heavy-flavor quarks or electroweak bosons (e.g., \(t\bar{t}\), \(W/Z\)), a lepton veto is applied. Events containing isolated electrons or muons are excluded, as they are often associated with background processes that do not align with the signal hypothesis of an invisible ALP produced in association with jets.

    \item \textbf{Azimuthal angle separation, $\Delta \phi(\text{jet}_{123}, M_{ET}) > 20^\circ$:} \\
    To ensure that the missing transverse energy is not aligned with the leading jets, an azimuthal angle separation 
    between the direction of the three leading jets (\({\rm jet}_{123}\)) and \(M_{ET}\) is enforced. 
    A separation of larger than \(20^\circ\) is required to reduce backgrounds, such as QCD multijet events, 
    where the \(M_{ET}\) arises from mis-measured jet energies rather than genuine invisible particles.

    \item \textbf{At least two $b$-tagged jets:} \\
    Events are required to contain at least two \(b\)-tagged jets, as the decay of the Higgs boson to two \(b\)-quarks (\(h \to b\bar{b}\)) is a key signature of the signal. The requirement of two \(b\)-tagged jets improves the signal-to-background ratio by reducing contributions from processes not involving \(b\)-quarks, such as light-quark or gluon-initiated jets.

    \item \textbf{Higgs transverse momentum selection, $P_{T_h}$:}
    \begin{itemize}
        \item $P_{T_h} > 100$ GeV if $M_{ET} < 350$ GeV.
        \item $P_{T_h} > 300$ GeV if $M_{ET} > 350$ GeV.
    \end{itemize}
    To ensure the transverse momentum (\(P_T\)) of the reconstructed Higgs boson is sufficiently large relative to the missing transverse energy, the transverse momentum selection depends on the \(M_{ET}\) value. A higher \(P_T\) threshold is applied for events with higher \(M_{ET}\), ensuring that the momentum balance between the Higgs and the invisible particles is consistent with the signal hypothesis.

    \item \textbf{Transverse mass of $b$-jets, $m_T^b$:} \\
    The transverse mass \(m_T^b\) of the \(b\)-jets is a key discriminant used to reduce background contributions from \(t\bar{t}\) events, where the top quark decays into a \(b\)-jet and a \(W\) boson. The transverse mass is calculated as:
    \[
    m_{T}^{b,\text{min/max}} = \sqrt{2 p_{T}^{b,\text{min/max}} E_{\text{miss}}^T \left(1 - \cos \Delta\phi(p_{T}^{b,\text{min/max}}, E_{\text{miss}}^T)\right)}
    \]
    where \(\Delta \phi\) is the azimuthal angle between the \(b\)-jet momentum and the missing transverse energy. Two cuts are applied:
    \begin{itemize}
        \item $m_T^{b,\min} > 170$ GeV.
        \item $m_T^{b,\max} > 200$ GeV.
    \end{itemize}
    These thresholds help suppress the \(t\bar{t}\) background by exploiting the kinematic differences between signal and background processes.

    \item \textbf{Number of jets:}
    \begin{itemize}
        \item $N_{\text{jets}} \leq 4$ for events with 2 b-tags.
        \item $N_{\text{jets}} \leq 5$ for events with $\geq$ 3 b-tags.
    \end{itemize}
    The number of jets in the event is restricted to further suppress background events, particularly from QCD multijet production, 
    which typically results in a higher jet multiplicity. 
    
    \item \textbf{Invariant mass of $b$-jet pair, $50$ GeV $< m_{b\bar{b}} < 280$ GeV:} \\
    The invariant mass of the two leading \(b\)-jets is required to fall within the range expected for 
    Higgs decay to \(b\)-quarks. The mass window of \(50~\text{GeV} < m_{b\bar{b}} < 280~\text{GeV}\) ensures 
    that events are selected around the Higgs mass peak while allowing for detector resolution effects. 
    This requirement significantly reduces background contributions from non-Higgs processes 
    while retaining the majority of the signal.
\end{itemize}

\begin{table}[t!]
\centering
\begin{ruledtabular}
\begin{tabular}{lccc}
    & $pp\to t\bar{t}$ & $pp\to W+HF$ & $pp\to Z+HF$ \\
\hline
$\sigma(\sqrt{s}=13~\rm TeV)$ (pb) & 833.9  & 361.1  & 1014  \\
$\sigma(\sqrt{s}=14~\rm TeV)$ (pb) & 985.7  & 395  & 1143  \\

\hline
\textbf{2-btag} & & & \\
$\rm N_b(\sqrt{s}=13~\rm TeV):150 \leq M_{ET}<200$ GeV & 4680  & 1590  & 6470  \\
$\rm N_b(\sqrt{s}=13~\rm TeV):200 \leq M_{ET}<350$ GeV & 3280 & 1760 & 7200 \\
$\rm N_b(\sqrt{s}=13~\rm TeV):350 \leq M_{ET}<500$ GeV & 76 & 106 & 507 \\
$\rm N_b(\sqrt{s}=13~\rm TeV):500 \leq M_{ET}<750$ GeV & 11.4  & 25  & 94  \\
$\rm N_b(\sqrt{s}=13~\rm TeV):M_{ET}>750$ GeV & 0.38 & 3.1 & 9.2 \\

$\rm N_b^{Rescaled}(\sqrt{s}=14~\rm TeV):150 \leq M_{ET}<200$ GeV & 5532 & 1739 & 7293 \\
$\rm N_b^{Rescaled}(\sqrt{s}=14~\rm TeV):200 \leq M_{ET}<350$ GeV & 3877 & 1925 & 8116 \\
$\rm N_b^{Rescaled}(\sqrt{s}=14~\rm TeV):350 \leq M_{ET}<500$ GeV & 90 & 116 & 571 \\
$\rm N_b^{Rescaled}(\sqrt{s}=14~\rm TeV):500 \leq M_{ET}<750$ GeV & 13.47 & 27.34 & 105 \\
$\rm N_b^{Rescaled}(\sqrt{s}=14~\rm TeV):M_{ET}>750$ GeV& 0.45 & 3.39 & 10.37 \\

\hline
\textbf{3-btag} & & & \\
$\rm N_b(\sqrt{s}=13~\rm TeV):150 \leq M_{ET}<200$ GeV & 276  & 21  & 102  \\
$\rm N_b(\sqrt{s}=13~\rm TeV):200 \leq M_{ET}<350$ GeV & 252 & 47& 278 \\
$\rm N_b(\sqrt{s}=13~\rm TeV):350 \leq M_{ET}<500$ GeV & 5.1 & 4.2 & 26.4 \\
$\rm N_b(\sqrt{s}=13~\rm TeV):M_{ET}>500$ GeV & 17.9 & 2.4 & 15.6 \\

$\rm N_b^{Rescaled}(\sqrt{s}=14~\rm TeV):150 \leq M_{ET}<200$ GeV & 326.24 & 22.95 & 114 \\
$\rm N_b^{Rescaled}(\sqrt{s}=14~\rm TeV):200 \leq M_{ET}<350$ GeV & 297.87 & 51.41 & 313 \\
$\rm N_b^{Rescaled}(\sqrt{s}=14~\rm TeV):350 \leq M_{ET}<500$ GeV & 6.02 & 4.59 & 29.75 \\
$\rm N_b^{Rescaled}(\sqrt{s}=14~\rm TeV):M_{ET}>500$ GeV & 21.15 & 2.62 & 17.58 \\

\end{tabular}
\end{ruledtabular}
\caption{Potential backgrounds, including $t\bar t$, $W + HF$ and $Z+HF$ 
where "HF" stands for heavy flavors, for the signal process 
$h\to b \bar{b}$ plus missing energies at the LHC with $\sqrt{s}= 13$ and 14 TeV.
Both 2 $b$-tag and 3 $b$-tag are shown. The numbers of background events are for $\sqrt{s}=13$ TeV 
and integrated luminosity ${\cal L}=139\, {\rm fb}^{-1}$ are adopted from the ATLAS Experimental paper \cite{ATLAS:2021shl}. 
Those for $\sqrt{s}= 14$ TeV with the same integrated luminosity are evaluated 
using the simplified rescaling in Eq.~(\ref{Eq.Bg}). 
Both the $\sqrt{s}=13$ and 14 TeV cross sections are calculated using \texttt{MadGraph5\_aMC@NLO}.}
\label{BGtab}
\end{table}
Major backgrounds in the signal regions are dominated by $t\bar{t}$ and $W/Z$ bosons produced with heavy-flavor jets. 
The $W/Z + \text{jets}$ background is categorized based on the jet flavor forming the Higgs candidate: 
if one or both jets are $b$-quarks, the event is classified as $W/Z + \text{HF}$, where HF refers to heavy-flavor. 
In the 2 $b$-tag resolved regions, the primary backgrounds are $t\bar{t}$ and $Z + \text{HF}$, with $Z + \text{HF}$ 
becoming more significant as $M_{ET}$ increases. In the 2 $b$-tag merged regions, $Z + \text{HF}$ dominates. 
Both the resolved and merged regions with $\geq 3$ $b$-tags are primarily affected by $t\bar{t}$, 
where the extra $b$-jet often comes from a mis-tagged hadronic $W$ decay. At high $M_{ET}$, $Z + \text{HF}$ 
backgrounds also become significant. The total number of background events in the 2 $b$-tag and 3 $b$-tag regions 
are provided in Table~4 and Table~5 of the ATLAS paper~\cite{ATLAS:2021shl}. 
The cut-flow table for the signal cross sections at $\sqrt{s}= 13$~TeV is shown in Table~\ref{cutflow13}. 
The signal selection criteria for three ALP masses, $m_a = 1$, $30$, and $60$~GeV, 
are presented in Table~\ref{cutflow13}. The invariant mass distribution of $m_{bb}$ for $b$-quarks 
at $\sqrt{s}=13$~TeV in different missing energy regimes is displayed in the left panel of FIG.~\ref{fig.mbb} 
(for ALP mass $m_a = 30$~GeV).

\begin{table}[t!]
\centering
\begin{ruledtabular}
\begin{tabular}{lccc}
\textbf{Cut} & \textbf{$m_a = 1$ GeV} & \textbf{$m_a = 30$ GeV} & \textbf{$m_a = 60$ GeV} \\
\hline
Total no. of events                               & 30144.23  & 22950.11  & 3185.93 \\
$M_{ET} > 150$ GeV                             & 13953.86  & 10494.80  & 1474.31 \\
Lepton veto                                & 13951.47  & 10493.88  & 1474.05 \\
Muon veto                                  & 13938.56  & 10486.71  & 1473.16 \\
Tauon veto                                 & 13055.50  & 9794.66   & 1377.31 \\
$\Delta \phi(jet_{123}, M_{ET}) > 20^\circ$ & 11941.38  & 8950.26   & 1263.31 \\
$M_{ET} < 500$ GeV                            & 11687.65  & 8768.98   & 1237.73 \\
$\geq 2$ b-tagged small-R jets             & 4405.05   & 3352.96   & 472.30  \\
$P_T^h$ selection                          & 3737.06   & 2868.71   & 399.00  \\
$m_T^{b,\min} > 170$ GeV                   & 1485.30   & 1137.36   & 159.30  \\
$m_T^{b,\max} > 200$ GeV                   & 933.26    & 715.89    & 101.06  \\
$N_{\text{jets}}$ selection                & 658.35    & 499.66    & 72.13   \\
$m_{b\bar{b}}$ mass selection              & 595.25    & 446.61    & 65.93   \\
\hline
\textbf{2-btag} & & & \\
$150 \leq M_{ET}<200$ GeV                    & 169.41    & 148.92    & 19.23   \\
$200 \leq M_{ET}<350$ GeV                    & 367.76    & 263.93    & 40.31   \\
$350 \leq M_{ET}<500$ GeV                    & 38.58     & 24.81     & 4.52    \\
\hline
\textbf{3-btag} & & & \\
$150 \leq M_{ET}<200$ GeV                    & 4.22      & 1.84      & 0.32    \\
$200 \leq M_{ET}<350$ GeV                    & 13.26     & 6.89      & 1.47    \\
$350 \leq M_{ET}<500$ GeV                    & 2.41      & 0.92      & 0.06    \\
\end{tabular}
\end{ruledtabular}
\caption{Cut-flow table for various values of $m_a$ in the resolved region $M_{ET}<500$ GeV for the HL-LHC ($\sqrt{s}=14$ TeV, 
$\mathcal{L}=3000~\mathrm{fb^{-1}}$). The coupling is fixed by setting $C_{aH}=1$ and $\Lambda = 1000$ GeV. 
The numbers of events are calculated by Eq.~(\ref{eq:Event_rate}).}
\label{cutflow14}
\end{table}

\subsubsection{ALP Signal and Backgrounds at HL-LHC} \label{sec.3a2}

For the generation of signal events at the LHC with $\sqrt{s}=14$~TeV and event selection, 
we followed the same formalism discussed in Sec.~\ref{sec.3a1}. 
However, the upper bounds on the model-independent cross sections from \cite{ATLAS:2021shl} 
cannot be used to estimate the ALP-Higgs coupling that one can achieve at the HL-LHC.
Instead, we applied a simple rescaling approach for the major background processes discussed in the ATLAS paper \cite{ATLAS:2021shl}:
\begin{equation}\label{Eq.Bg}
N^{\text{Rescaled}}_b(\sqrt{s}=14 ~\text{TeV}) = N_b(\sqrt{s}=13 ~\text{TeV}) \times \frac{\sigma(\sqrt{s}=14 ~\text{TeV})}{\sigma(\sqrt{s}=13 ~\text{TeV})}
\end{equation}
Using this rescaling relation, we calculated the number of background events at the 14~TeV LHC across different missing energy regions. 
In Eq.~(\ref{Eq.Bg}), $N_b(\sqrt{s}=13 ~\text{TeV})$ represents the number of background events for the major 
processes ($t\bar{t}$, $Z+HF$, $W+HF$) reported in the ATLAS paper \cite{ATLAS:2021shl}. 
The cross sections, $\sigma(\sqrt{s}=14 ~\text{TeV})$ and $\sigma(\sqrt{s}=13 ~\text{TeV})$, for the respective background processes 
are evaluated using \texttt{MadGraph5\_aMC@NLO} without any parton-level cuts. 
Table~\ref{BGtab} compares the number of background events reported in the ATLAS paper \cite{ATLAS:2021shl} 
with those evaluated using the rescaling formula in Eq.~(\ref{Eq.Bg}). Note that the background event rates for $\sqrt{s}=14$~TeV shown in Table~\ref{BGtab} 
are for the same luminosity ${\cal L}=139\,{\rm fb}^{-1}$. Later, when we estimate the sensitivities at
the HL-LHC, we rescale the background event rates by $3000 / 139$.

The cut flow table for the LHC with $\sqrt{s}=14$~TeV and ${\cal L} =3000\,{\rm fb}^{-1}$ is presented 
in Table~\ref{cutflow14} for three ALP masses: $m_a=1$, $30$, and $60$~GeV. 
When comparing the cut flow in Table~\ref{cutflow14} for $\sqrt{s}=14$~TeV to that at $\sqrt{s}=13$~TeV 
in Table~\ref{cutflow13}, it is evident that the event selection follows similar pattern in both cases. 
The distribution of the invariant mass $m_{bb}$ at $\sqrt{s}=14$~TeV, across various missing energy ranges, is presented 
in the right panel of FIG.~\ref{fig.mbb} (for an ALP mass of $m_a = 30$GeV). 
%
%
It is evident from Tables \ref{cutflow13} and \ref{cutflow14} that the missing energy bins with 3-btags 
excessively remove more signal events, thus would result in weaker sensitivities.
Therefore, we will proceed with our calculations using the 2-btag case.


\subsection{Sterile Neutrino Signal and Backgrounds} \label{sec.3b}

\begin{figure}[h!]
    \centering
    \begin{subfigure}[b]{0.49\textwidth}
        \centering
        \includegraphics[width=\textwidth]{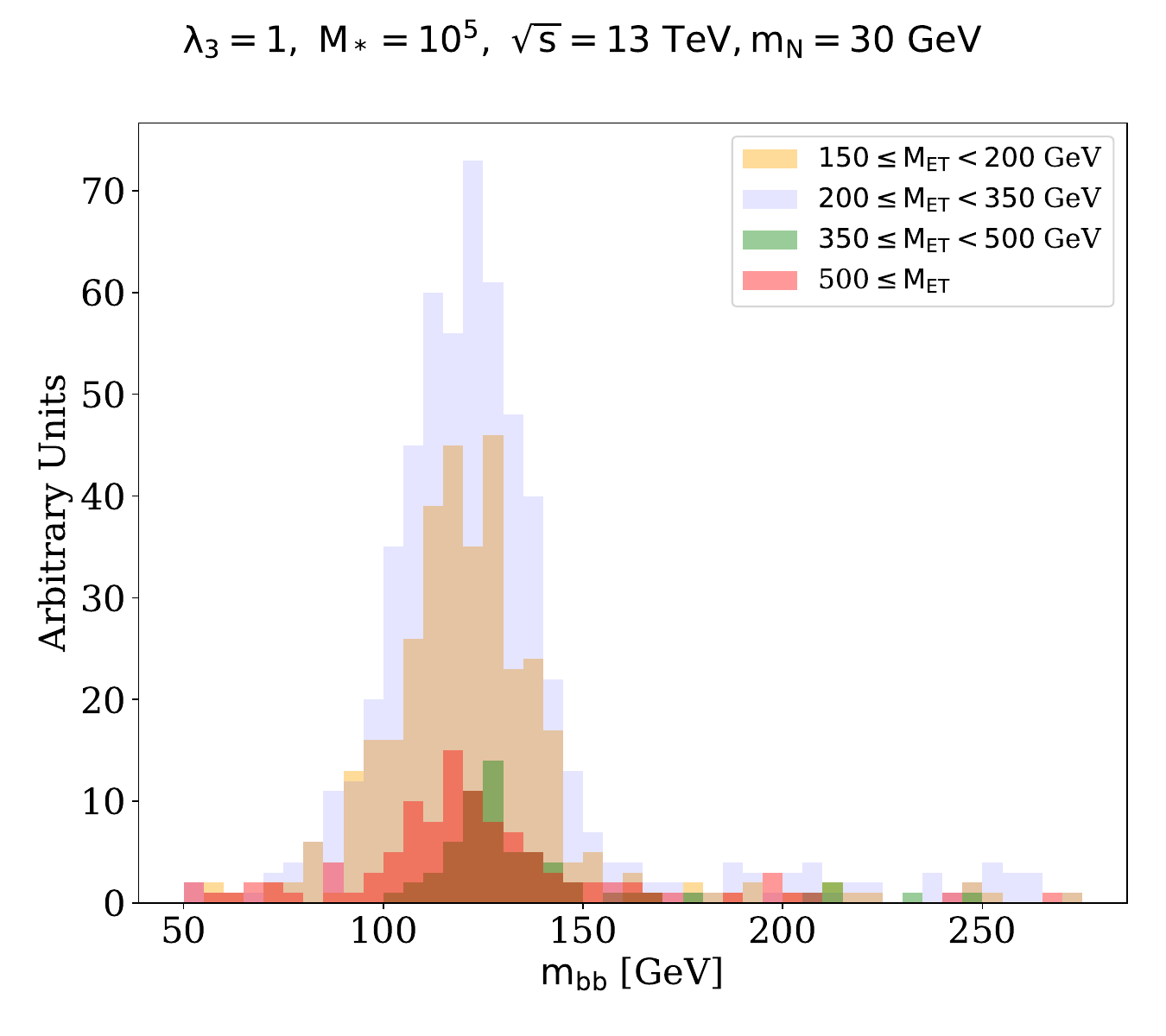}
        \label{fig.dis_13TeV_S}
    \end{subfigure}
    \hfill
    \begin{subfigure}[b]{0.49\textwidth}
        \centering
        \includegraphics[width=\textwidth]{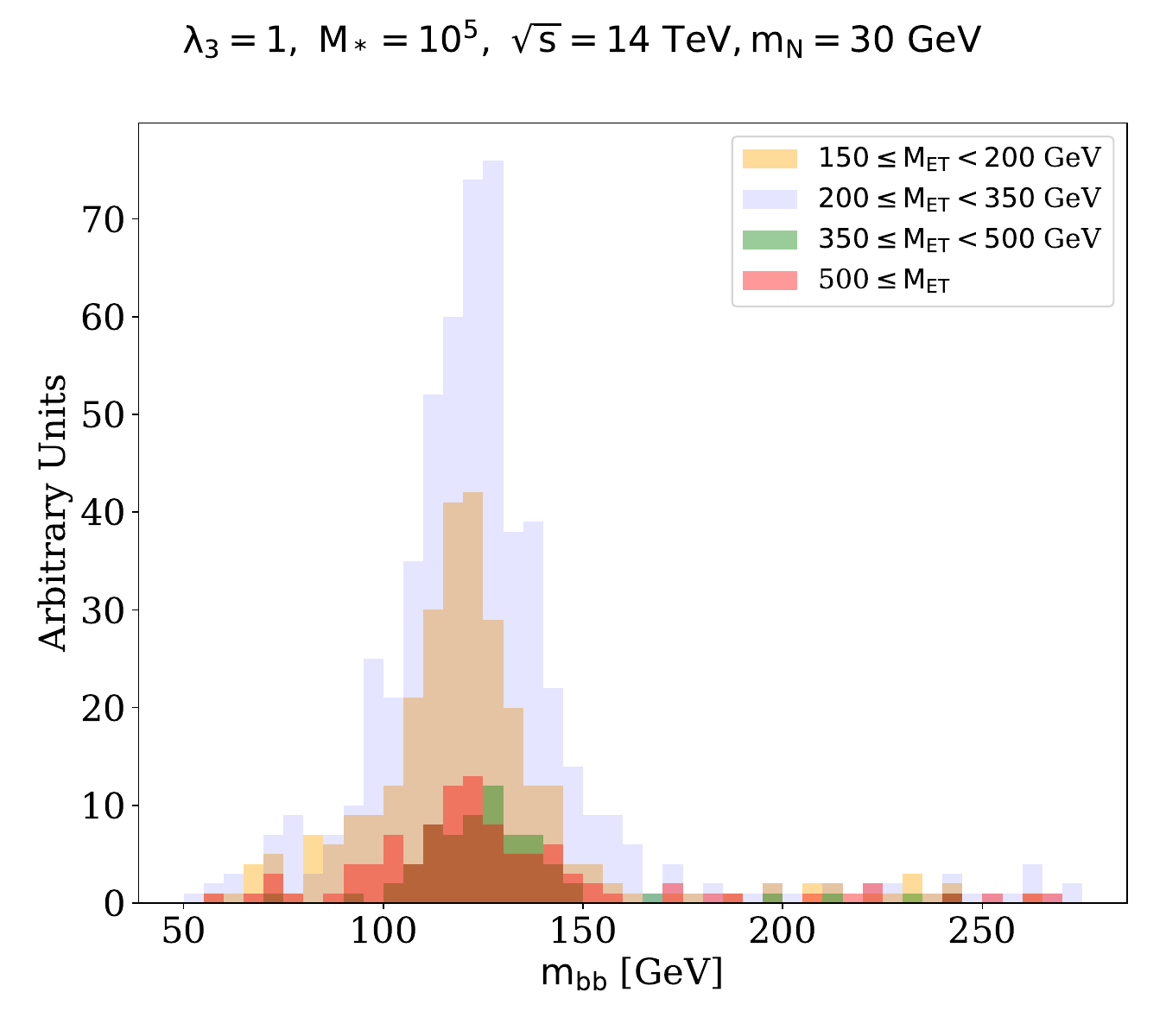}
        \label{fig.dis_14TeV_S}
    \end{subfigure}
    \caption{Invariant mass $m_{bb}$ distributions for the $b$-quark pair in the process
    $pp \to h N N \,;\, h \to b \bar b$ at the LHC with (a) $\sqrt{s}=13$ TeV and (b) $\sqrt{s}=14$ TeV in various missing energy regimes.
    }
    \label{fig.mbb_Ster}
\end{figure}

Sterile neutrino events are simulated following the formalism discussed in Sec.~\ref{sec.3a}. 
We present the signal event analysis for both the LHC (\( \sqrt{s} = 13~\text{TeV}, \mathcal{L} = 139~\text{fb}^{-1} \)) 
and the HL-LHC (\( \sqrt{s} = 14~\text{TeV}, \mathcal{L} = 3000~\text{fb}^{-1} \)) with three different sterile neutrino masses:
\( m_N = 1, 30, \) and \( 60 \) GeV. The invariant mass distribution \( m_{bb} \) for \( b \)-quarks at 13~TeV, 
across various missing energy regimes, is shown in the left panel of 
FIG.~\ref{fig.mbb_Ster} (for a sterile neutrino mass \( m_N = 30 \)~GeV).

As in the previous subsection on ALPs, we followed the ATLAS paper for the signal event analysis. 
Table~\ref{cutflow13_St} presents the cut flow for the sterile neutrino signal event analysis at the LHC (\(\sqrt{s} = 13\) TeV,
\(\mathcal{L} = 139\) fb\(^{-1}\)). In this simulation, we generated \(5 \times 10^5\) signal events 
sing \texttt{MadGraph5\_aMC@NLO}. Similar to the case of ALPs, we present the results of sterile neutrinos 
for the resolved region (\(M_{ET} < 500~\text{GeV}\)) in the main text, while the cut flow table for 
the merged region (\(M_{ET} > 500~\text{GeV}\)) is presented in Appendix~\ref{App2}. 
We fix the coupling at $\rm \lambda_3=1$ and $\rm M_* = 10^5$ GeV, such that the
event rates scale as $(\lambda_3 / M_*)^2$. 
The cut flow follows a similar pattern to that of the ALPs.
 
\begin{table}[t!]
\centering
\begin{ruledtabular}
\begin{tabular}{lccc}
\textbf{Cut} & \textbf{$m_N = 1$ GeV} & \textbf{$m_N = 30$ GeV} & \textbf{$m_N = 60$ GeV} \\
\hline
Total events                               & 79.177 & 53.731 & 1.764 \\
$M_{ET} > 150$ GeV                            & 35.850 & 23.835 & 0.798 \\
Lepton veto                                & 35.845 & 23.833 & 0.798 \\
Muon veto                                  & 35.829 & 23.811 & 0.797 \\
Tauon veto                                 & 33.611 & 22.326 & 0.747 \\
$\Delta \phi(jet_{123}, M_{ET}) > 20^\circ$ & 30.884 & 20.451 & 0.686 \\
$M_{ET} < 500$ GeV                            & 30.286 & 20.060 & 0.670 \\
$\geq 2$ b-tagged small-R jets             & 11.547 & 7.582 & 0.251 \\
$P_T^h$ selection                          & 9.884  & 6.463 & 0.213 \\
$m_T^{b,\min} > 170$ GeV                   & 3.885  & 2.556 & 0.085 \\
$m_T^{b,\max} > 200$ GeV                   & 2.484  & 1.611 & 0.054 \\
$N_{\text{jets}}$ selection                & 1.823  & 1.161 & 0.039 \\
$m_{b\bar{b}}$ mass selection              & 1.661  & 1.053 & 0.036 \\
\hline
\textbf{2-btag} & & & \\
$150 \leq M_{ET}<200$ GeV                    & 0.524  & 0.363 & 0.011 \\
$200 \leq M_{ET}<350$ GeV                    & 0.986  & 0.596 & 0.021 \\
$350 \leq M_{ET}<500$ GeV                   & 0.120  & 0.066 & 0.003 \\
\hline
\textbf{3-btag} & & & \\
$150 \leq M_{ET}<200$ GeV                    & 0.013  & 0.005 & 0.0003 \\
$200 \leq M_{ET}<350$ GeV                     & 0.016  & 0.023 & 0.0007 \\
$350 \leq M_{ET}<500$ GeV                    & 0.002  & 0.001 & 0.0002 \\
\end{tabular}
\end{ruledtabular}
\caption{Cut-flow table for various values of $m_N$ in the resolved region $M_{ET}<500$ GeV for the LHC ($\sqrt{s}=13$ TeV, $\mathcal{L}=139~\rm fb^{-1}$). The coupling is fixed by setting 
$\rm \lambda_3=1$ and $\rm M_* = 10^5$ GeV. The numbers of events are calculated by Eq.~(\ref{eq:Event_rate}).}
\label{cutflow13_St}
\end{table}

\subsubsection{Sterile Neutrino signal and Backgrounds at the HL-LHC} \label{sec.3b1}

This section presents the production of sterile neutrinos at the HL-LHC (\(\sqrt{s} = 14\) TeV, \(\mathcal{L} = 3000\) fb\(^{-1}\)). As in the earlier ALP analysis, we adopt the signal event analysis framework from the ATLAS paper. Table~\ref{cutflow14_St} shows the sterile neutrino signal event analysis at the HL-LHC (\(\sqrt{s} = 14\) TeV, \(\mathcal{L} = 3000\) fb\(^{-1}\)).

For this simulation, we generated \(5 \times 10^5\) signal events using \texttt{MadGraph5\_aMC@NLO}. 
Similar to the ALPs, we also focus on the resolved region (\(M_{ET} < 500~\text{GeV}\)) in the main text, 
while the cut flow table for the merged region (\(M_{ET} > 500~\text{GeV}\)) is listed in Appendix~\ref{App2}. 
In this setup, the coupling is fixed at \(\lambda_3 = 1\) and \(M_* = 10^5\) GeV. 
The event progression in the cut flow table mirrors the structure of the ALP analysis. 
The rescaled background event rates for the HL-LHC with \(\sqrt{s} = 14\) TeV are given in Table.~\ref{BGtab}.

\begin{table}[t!]
\centering
\begin{ruledtabular}
\begin{tabular}{lccc}
\textbf{Cut} & \textbf{$m_N = 1$ GeV} & \textbf{$m_N = 30$ GeV} & \textbf{$m_N = 60$ GeV} \\
\hline
Total events                               & 2020.26  & 1370.99  & 45.01 \\
$M_{ET} > 150$                             & 929.88   & 618.80   & 20.43 \\
Lepton veto                                & 929.72   & 618.56   & 20.42 \\
Muon veto                                  & 929.09   & 617.95   & 20.42 \\
Tauon veto                                 & 870.36   & 579.08   & 19.10 \\
$\Delta \phi(jet_{123}, M_{ET}) > 20^\circ$ & 795.95  & 530.30   & 17.46 \\
$M_{ET} < 500$                             & 778.35  & 519.36   & 17.08 \\
$\geq 2$ b-tagged small-R jets             & 299.65  & 193.81   & 6.38 \\
$P_T^h$ selection                          & 254.83  & 165.94   & 5.47 \\
$m_T^{b,\min} > 170$ GeV                   & 103.16  & 66.05    & 2.20 \\
$m_T^{b,\max} > 200$ GeV                   & 63.01    & 40.17    & 1.37 \\
$N_{\text{jets}}$ selection                & 45.08    & 28.33    & 0.98 \\
$m_{b\bar{b}}$ mass selection              & 40.60    & 25.27    & 0.89 \\
\hline
\textbf{2-btag} & & & \\
$150 \leq M_{ET}<200$                     & 12.32    & 7.78     & 0.26 \\
$200 \leq M_{ET}<350$                     & 24.39    & 15.03    & 0.55 \\
$350 \leq M_{ET}<500$                     & 3.11     & 1.92     & 0.07 \\
\hline
\textbf{3-btag} & & & \\
$150 \leq M_{ET}<200$                     & 0.12     & 0.22     & 0.004 \\
$200 \leq M_{ET}<350$                     & 0.65     & 0.30     & 0.015 \\
$350 \leq M_{ET}<500$                     & 0.04     & 0.03     & 0.004 \\
\end{tabular}
\end{ruledtabular}
\caption{Cut-flow table for various values of $m_N$ in the resolved region $M_{ET}<500$ GeV for the HL-LHC ($\sqrt{s}=14$ TeV, $\mathcal{L}=3000~\mathrm{fb^{-1}}$). The coupling is fixed by setting 
$\lambda_3=1$ and $ M_* = 10^5 $ GeV. The numbers of events are calculated by Eq.~(\ref{eq:Event_rate}).}
\label{cutflow14_St}
\end{table}

%% file: subtex/04_result.tex
\section{Results}\label{sec.4}

\subsection{Results on ALP} \label{sec.4a}

To evaluate the exclusion regions at 95\% confidence level (C.L.) for the ALP-Higgs coupling \(\frac{C_{aH}}{\Lambda^2}\) at the LHC with \(\sqrt{s} = 13\) TeV and \(\mathcal{L} = 139~\text{fb}^{-1}\), we closely follow the experimental cuts outlined in the ATLAS paper~\cite{ATLAS:2021shl}, as mentioned in the previous section, in order to directly apply their model-independent upper limits on the visible cross-section across different signal regions. For the HL-LHC case with \(\sqrt{s} = 14\) TeV and \(\mathcal{L} = 3000~\text{fb}^{-1}\), we use the expression for significance \(Z\), given in Eq.~(\ref{eq.sig}), as a function of signal (\(N_s\)) and background (\(N_b\)) events, incorporating systematic uncertainty \(\sigma\).

After applying the event selection criteria outlined in the previous section, the number of signal events is comparable to, or even greater than, the background events. This makes it important to evaluate the signal significance and establish limits on the cutoff scale, $\frac{C_{aH}}{\Lambda^2}$. The relationship between the number of ALP signal events, $N_s$, and the cutoff scale, $\frac{C_{aH}}{\Lambda^2}$, follows the proportionality:
\begin{equation}
    N_s \propto (\frac{C_{aH}}{\Lambda^2})^{2}.
\end{equation}
Thus, the scale $\frac{C_{aH}}{\Lambda^2}$ can be adjusted to align with the expected number of signal events, $N_s$. The signal significance is calculated using the expression 
\cite{Cousins:2007yta}
\begin{equation}\label{eq.sig}
    Z = \sqrt{2\left[(N_s+N_b)\ln\left(\frac{(N_s+N_b)(N_b+\sigma^2)}{{N_b}^2+(N_s+N_b)\sigma^2}\right) - \frac{N_b^2}{\sigma^2}\ln\left(1+\frac{\sigma^2 N_s}{N_b(N_b+\sigma^2)}\right)\right]},
\end{equation}
where $N_s$ and $N_b$ represent the number of signal and background events, respectively, and $\sigma$ denotes the systematic uncertainty in the background estimation. In the analysis, $\sigma$ is considered as either zero or $0.2 N_b$. The $95\%$ confidence level (C.L.) sensitivity bounds on the ALP cutoff scale, $\frac{C_{aH}}{\Lambda^2}$, as a function of the ALP mass, $M_a$, are determined by requiring a significance $Z > 2$.

\begin{figure}[h!]
    \centering
    \includegraphics[width=\textwidth]{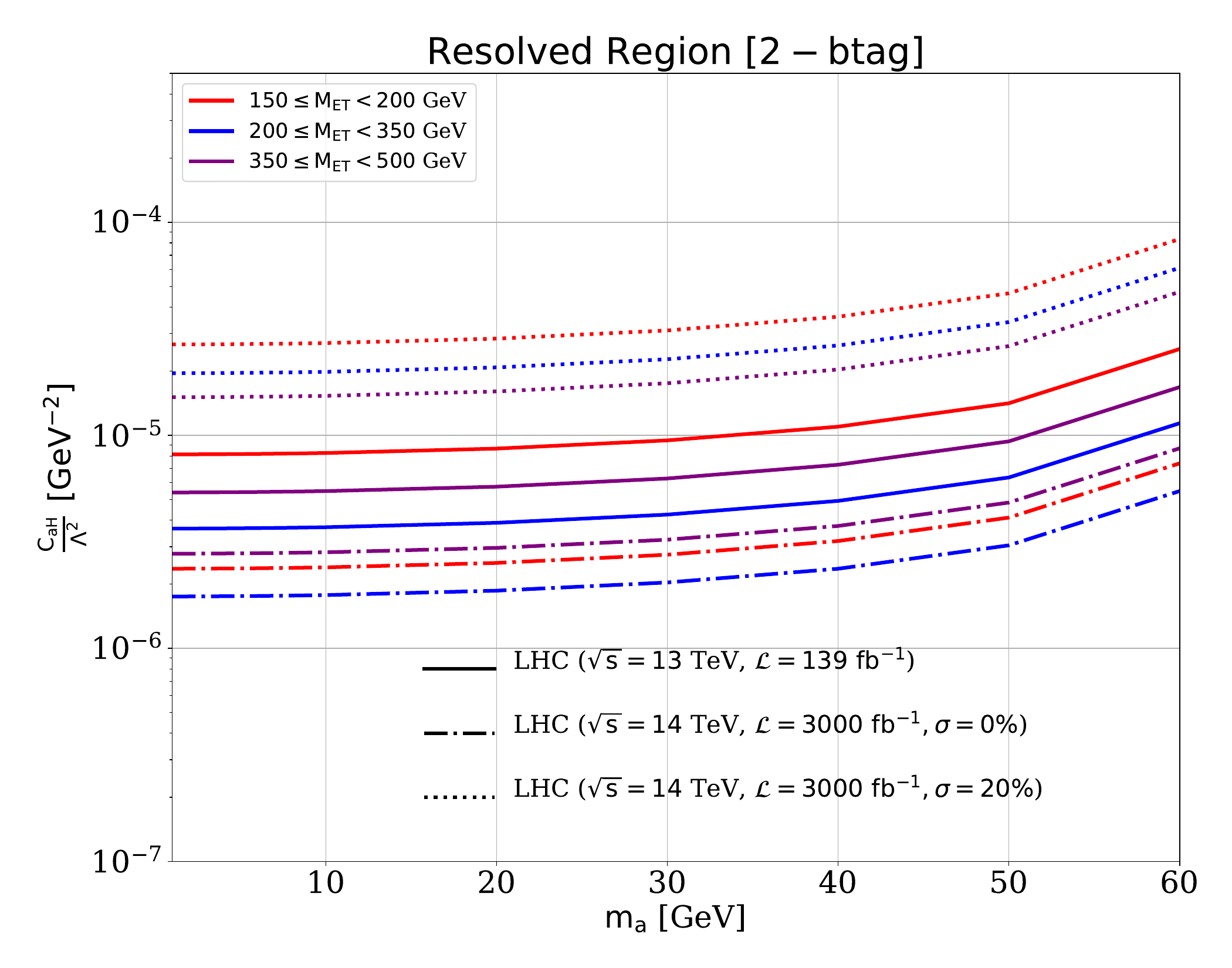}
    \caption{Exclusion regions (above each solid curve) at 95\% confidence level (C.L.) for the ALP-Higgs coupling  
$\frac{C_{aH}}{\Lambda^2}$ at the LHC with $\sqrt{s} = 13$ TeV and an integrated luminosity of  
$\mathcal{L} = 139~\mathrm{fb}^{-1}$. Sensitivity regions (above each dotted curve) at 95\% C.L. are shown  
for the HL-LHC with $\sqrt{s} = 14$ TeV and $\mathcal{L} = 3000~\mathrm{fb}^{-1}$ in the resolved region  
considering a 20\% systematic uncertainty. Additionally, sensitivity regions (above each dot-dashed curve)  
at 95\% C.L. for the HL-LHC ($\sqrt{s} = 14$ TeV, $\mathcal{L} = 3000~\mathrm{fb}^{-1}$) are presented  
in the resolved region without systematic uncertainty.
}
    \label{fig.Sen}
\end{figure}

FIG.~\ref{fig.Sen} presents the exclusion regions at 95\% C.L. for the ALP-Higgs coupling $\frac{C_{aH}}{\Lambda^2}$ 
at the LHC and HL-LHC. In the figure, solid lines depict the exclusion limits for the ALP-Higgs coupling 
at the LHC with $\sqrt{s} = 13$ TeV and an integrated luminosity of $\mathcal{L} = 139~\mathrm{fb}^{-1}$, 
while the dotted lines show the projected limits for the HL-LHC at $\sqrt{s} = 14$ TeV with $\mathcal{L} = 3000~\mathrm{fb}^{-1}$, 
including a 20\% systematic uncertainty, 
whereas the dot-dashed lines represent the 14 TeV HL-LHC projections ($\mathcal{L} = 3000~\mathrm{fb}^{-1}$) 
without accounting for systematic uncertainty. 

The exclusion plot clearly indicates that the 200–350 GeV missing energy range provides stronger constraints
on the ALP-Higgs coupling. On the other hand, at the 14 TeV LHC with 20\% systematic uncertainty, 
the ALP-Higgs coupling achieves stronger bounds in the $350–500$ GeV missing energy range than in any other ranges. 
This is primarily due to smaller number of background events in this range; with the addition of a 20\% systematic uncertainty, 
this region achieves a higher significance compared to the others. 
This behavior can be inferred from the number of signal events in each missing energy range, 
as shown by the invariant mass distribution of \(b\)-quarks in FIG.~\ref{fig.mbb}. 
As expected, the HL-LHC with more than 20 times higher luminosity would yield significantly more stringent bounds on the ALP-Higgs coupling 
than the 13 TeV LHC data. Also, When comparing the exclusion plots in the resolved region (FIG.~\ref{fig.Sen}) 
to those in the merged region (FIG.~\ref{fig.Sen_Merge}) provided in Appendix~\ref{App1}, 
it is evident that the exclusion curves in the resolved region offer tighter constraints on the ALP-Higgs coupling. The LHC 14 TeV results with 20\% systematic uncertainty is less sensitive compared to 13 TeV results.

\subsection{Results on Sterile Neutrino}\label{sec.4b}

To determine the exclusion limits at the 95\% C.L. for the sterile neutrino-Higgs coupling \(\frac{\lambda_3}{M_*}\) 
at the LHC with \(\sqrt{s} = 13\) TeV and \(\mathcal{L} = 139~\text{fb}^{-1}\), 
we apply the experimental cuts specified in the ATLAS study~\cite{ATLAS:2021shl} as referenced previously. 
This enables us to directly utilize their model-independent upper bounds on the visible cross-section across various 
signal regions. For the HL-LHC scenario with \(\sqrt{s} = 14\) TeV and \(\mathcal{L} = 3000~\text{fb}^{-1}\), 
we employ the significance expression \(Z\) from Eq.~(\ref{eq.sig}), which depends on the signal (\(N_s\)) 
and background (\(N_b\)) event rates, alongside the systematic uncertainty \(\sigma\).

Following the event selection criteria described earlier, the number of signal events approaches or exceeds that of the background events, 
making it essential to compute the signal significance and set constraints on the coupling scale, 
\(\frac{\lambda_3}{M_*}\). The number of sterile neutrino signal events, \(N_s\), scales as:
\begin{equation}
    N_s \propto (\frac{\lambda_3}{M_*})^2.
\end{equation}
Consequently, the value of \(\frac{\lambda_3}{M_*}\) can be adjusted to achieve the expected signal event rates \(N_s\).

\begin{figure}[h!]
    \centering
    \includegraphics[width=\textwidth]{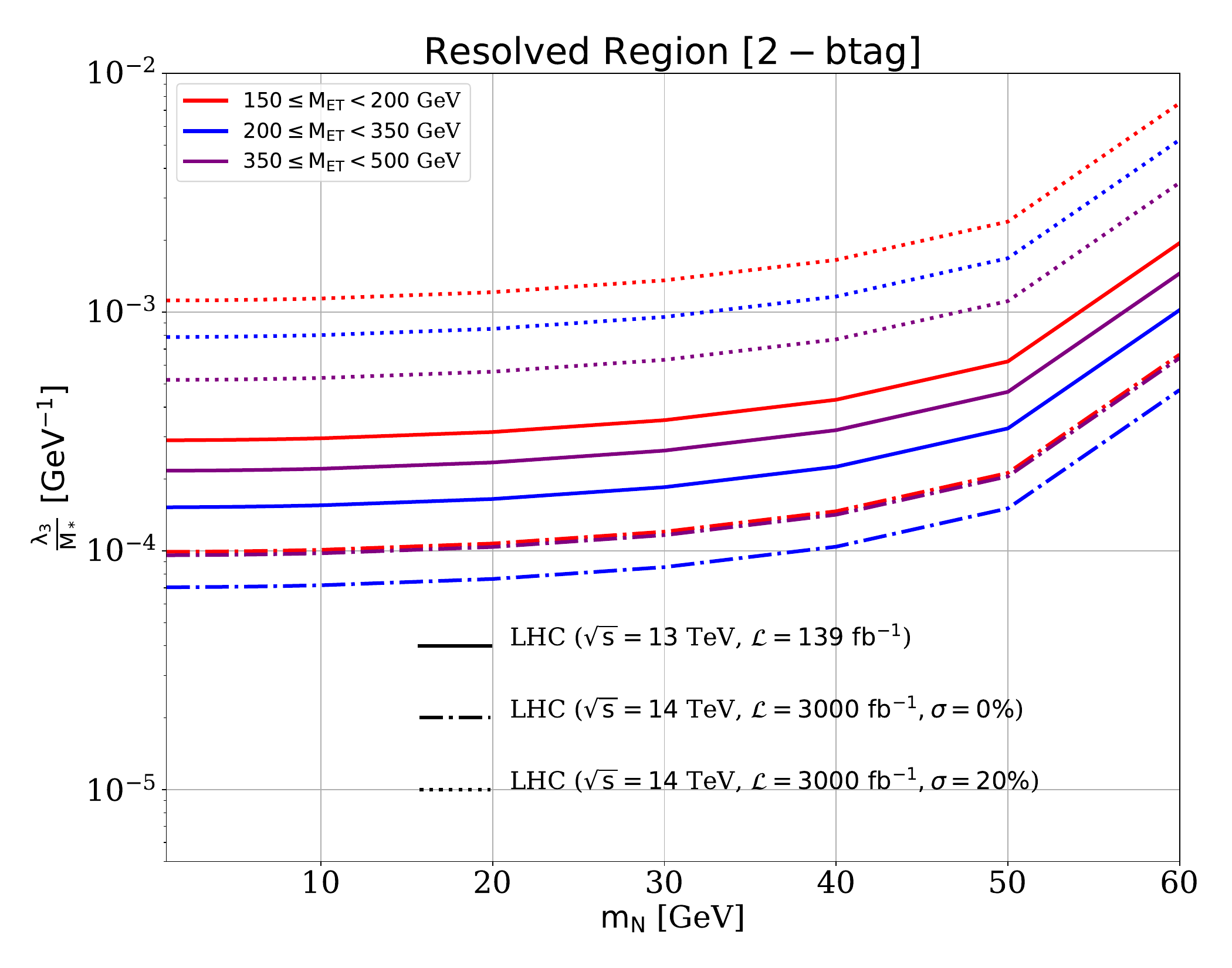}
    \caption{Exclusion regions (above each solid curve) at 95\% confidence level (C.L.) for the sterile neutrino-Higgs coupling  
$\frac{\lambda_3}{M_*}$ at the LHC with $\sqrt{s} = 13$ TeV and an integrated luminosity of  
$\mathcal{L} = 139~\mathrm{fb}^{-1}$. Sensitivity regions (above each dotted curve) at 95\% C.L. are shown  
for the HL-LHC with $\sqrt{s} = 14$ TeV and $\mathcal{L} = 3000~\mathrm{fb}^{-1}$ in the resolved region  
considering a 20\% systematic uncertainty. Additionally, sensitivity regions (above each dot-dashed curve)  
at 95\% C.L. for the HL-LHC ($\sqrt{s} = 14$ TeV, $\mathcal{L} = 3000~\mathrm{fb}^{-1}$) are presented  
in the resolved region without systematic uncertainty.
    }
    \label{fig.Sen_Sterile_Resolve}
\end{figure}

FIG.~\ref{fig.Sen_Sterile_Resolve} shows the exclusion regions at the 95\%  C.L. for the sterile 
neutrino-Higgs coupling \(\frac{\lambda_3}{M_*}\) at both the LHC and HL-LHC. In the figure, 
the solid lines represent the exclusion limits for the sterile neutrino-Higgs coupling at the LHC with \(\sqrt{s} = 13\) TeV 
and an integrated luminosity of \(\mathcal{L} = 139~\mathrm{fb}^{-1}\), 
while the dotted lines depict the projected limits for the HL-LHC at \(\sqrt{s} = 14\) TeV with \(\mathcal{L} = 3000~\mathrm{fb}^{-1}\), 
including a 20\% systematic uncertainty, and the dot-dashed lines represent the HL-LHC projections 
at 14 TeV (\(\mathcal{L} = 3000~\mathrm{fb}^{-1}\)) without considering systematic uncertainty.

The exclusion plot demonstrates that the missing energy range of 200–350 GeV provides a stronger 
constraint on the sterile neutrino-Higgs coupling. This pattern is evident from the number of signal events
within each missing energy interval, as illustrated by the invariant mass distribution of 
b-quarks in FIG.~\ref{fig.mbb_Ster}. As anticipated, the projections at the HL-LHC yield more stringent bounds 
on the sterile neutrino-Higgs coupling than the 13 TeV LHC data. A
dditionally, comparing the exclusion plots for the resolved region (FIG.~\ref{fig.Sen_Sterile_Resolve}) 
with those for the merged region (FIG.~\ref{fig.Sen_Sterile_Merge}) in Appendix~\ref{App2}, 
it is clear that the resolved region exclusion curves provide tighter constraints on the sterile neutrino-Higgs coupling.

%% file: subtex/05_Conclusion.tex
\section{Conclusions}\label{sec.5}

In this study, we have explored the interactions of the SM Higgs boson with two distinct hidden sectors via higher-dimensional operators, 
namely, (i) axion-like particles (ALP) and (ii) sterile neutrinos. 
The SM Higgs interaction with ALPs begins at the dimension-six level, 
represented by the Lagrangian term \( L = \frac{C_{aH}}{\Lambda^2} (\partial_\mu a) (\partial^\mu a) \). 
In contrast, the sterile neutrino-Higgs interaction starts at the dimension-five level, represented by \( L = \frac{\lambda_3}{M_*} H^{\dag} H N N \). 
To investigate these interactions, we analyzed the production of ALPs (\( pp \to h a a \)) and 
sterile neutrinos (\( pp \to h N N \)) at both the LHC and High Luminosity LHC (HL-LHC), 
focusing on the mono-Higgs plus missing energy signature where the Higgs subsequently decays 
as \( h \to b \bar{b} \). Both the ALPs and sterile neutrinos are considered stable particles 
with large missing energy, resulting in similar event topologies for both processes.


For the analysis at the LHC with \(\sqrt{s} = 13\) TeV and an integrated luminosity of \(\mathcal{L} = 139~\text{fb}^{-1}\), 
we closely followed the event selection criteria and experimental cuts provided in the ATLAS paper~\cite{ATLAS:2021shl}. 
This approach allowed us to directly apply the model-independent upper limits on the visible cross-section across across signal regions. 
The ATLAS paper~\cite{ATLAS:2021shl} also included a comprehensive signal and background analysis, 
which we used as a basis for this work. For the HL-LHC projections at \(\sqrt{s} = 14\) TeV and \(\mathcal{L} = 3000~\text{fb}^{-1}\), 
we rescaled the primary background events presented in the 13 TeV ATLAS study.

Our study establishes bounds on both the ALP-Higgs coupling \(\frac{C_{aH}}{\Lambda^2}\) as a function of the ALP mass \(m_a\) ranging from 1 to 60 GeV, 
and the sterile neutrino-Higgs coupling \(\frac{\lambda_3}{M_*}\) as a function of the sterile neutrino mass \(m_N\) 
within the same mass range, for various missing energy regimes at both LHC and HL-LHC settings. 
Notably, our results indicate that the missing transverse energy (\(M_{ET}\)) range of \(200 < M_{ET} \leq 350\) GeV 
provides the most stringent constraints on the model parameters.

In the search for both ALPs and sterile neutrinos at the LHC, the sensitivity results 
exhibit variation across different energy and luminosity regimes, 
highlighting distinct parameter dependencies for the respective particles. 
Here, we have presented the results from comparison studies at both the 13 TeV LHC with 
a luminosity of 139 fb\(^{-1}\) and the 14 TeV HL-LHC with an increased luminosity of 3000 fb\(^{-1}\).

In the resolved region (\(M_{ET} < 500\) GeV) at the 13 TeV LHC, for the missing transverse energy 
range \(200 \, \text{GeV} < M_{ET} \leq 350 \, \text{GeV}\), 
the sensitivity to the coupling parameter \(\frac{C_{aH}}{\Lambda^2}\) for ALPs reaches a peak of 
approximately \(3.8 \times 10^{-6} \,{\rm GeV}^{-2} \) at \(m_a = 1 \, \text{GeV}\), 
decreasing to \(1 \times 10^{-5} \,{\rm GeV}^{-2} \) at \(m_a = 60 \, \text{GeV}\). 
For sterile neutrinos, the sensitivity to \(\frac{\lambda_3}{M_*}\) peaks at 
around \(1.5 \times 10^{-4} \,{\rm GeV}^{-1} \) at \(m_N = 1 \, \text{GeV}\) and 
drops to \(10^{-3}  \,{\rm GeV}^{-1} \) at \(m_N = 60 \, \text{GeV}\), 
indicating a generally lower sensitivity relative to ALPs.

At the 14 TeV HL-LHC with a luminosity of 3000 fb\(^{-1}\), both particles 
show improved sensitivities across varying missing transverse energy regions and systematic uncertainties. 
In the \(350 \, \text{GeV} < M_{ET} \leq 500 \, \text{GeV}\) range with a 20\% systematic uncertainty, 
the sensitivity to \(\frac{C_{aH}}{\Lambda^2}\) for ALPs peaks at \(1.5 \times 10^{-5}  \,{\rm GeV}^{-2} \) 
for \(m_a = 1 \, \text{GeV}\), while for sterile neutrinos the sensitivity to 
\(\frac{\lambda_3}{M_*}\) reaches \(5 \times 10^{-4}  \,{\rm GeV}^{-1} \) at \(m_N = 1 \, \text{GeV}\). 
Without systematic uncertainties, ALPs exhibit an enhanced peak sensitivity of approximately 
\(1.8 \times 10^{-6}  \,{\rm GeV}^{-2}  \) in the \(200 \, \text{GeV} < M_{ET} \leq 350 \, \text{GeV}\) range, 
while sterile neutrinos reach \(7 \times 10^{-5}  \,{\rm GeV}^{-1}  \) under the same conditions.

In the merged region (\(M_{ET} > 500 \, \text{GeV}\)) at the HL-LHC, the 
differences between ALPs and sterile neutrinos remain evident. 
For ALPs, in the missing transverse energy range \(500 \, \text{GeV} \leq M_{ET} < 750 \, \text{GeV}\), 
the sensitivity to \(\frac{C_{aH}}{\Lambda^2}\) without systematic uncertainties reaches a 
peak value of about \(1 \times 10^{-6}  \,{\rm GeV}^{-2} \) at 
\(m_a = 1 \, \text{GeV}\), while the equivalent for sterile neutrinos 
is \(6 \times 10^{-5}  \,{\rm GeV}^{-1}  \) for \(\frac{\lambda_3}{M_*}\) at \(m_N = 1 \, \text{GeV}\).

\par For events with three b-tagged jets in the high missing transverse energy 
range (\(M_{ET} > 500 \, \text{GeV}\)), ALPs reach a sensitivity of approximately 
\(3.5 \times 10^{-6}  \,{\rm GeV}^{-2} \) at \(m_a = 1 \, \text{GeV}\), 
whereas sterile neutrinos attain \(1.8 \times 10^{-4}  \,{\rm GeV}^{-1} \) for \(m_N = 1 \, \text{GeV}\).
As \(m_a\) or \(m_N\) increases to 60 GeV, the sensitivity decreases for both particles, 
reaching \(1 \times 10^{-5}  \,{\rm GeV}^{-2}  \) for ALPs and \(1 \times 10^{-3}  \,{\rm GeV}^{-1} \) 
for sterile neutrinos, indicating a comparatively higher sensitivity for ALPs at smaller coupling values 
across the tested ranges.

Note that we did not turn on any dim-5 operators so that the ALP is stable on the
collider scale. On the other hand, if we turn on the dim-5 operators even on a very small level 
(see the calculation in II.A.2), the ALP would decay into a pair
of photons, gluons, or fermions. The final state would be different from $h aa\to (b\bar b) + \not E_T$,
instead, it would be  $h aa\to (b\bar b) + (\gamma\gamma/gg/f\bar f) (\gamma\gamma/gg/f\bar f)$. The new
final states would be more complex but could potentially be more interesting.

In addition, note that the operator $(\partial_\mu a)(\partial^\mu a) \phi^\dagger \phi$ can give rise 
to $H \to aa$, i.e., invisible Higgs decay. Similarly, the operator $H^\dagger H N N$ also gives
invisible decay $H \to NN$. Thus, one can use the upper limit on the invisible decay width of the
Higgs boson $B(H \to \; {\rm invisible}) < 0.16$ \cite {CMS:2022dwd} to constrain on the couplings.
We obtain $\frac{C_{ah}}{\Lambda^2} < 8.1 \times 10^{-7}~\text{GeV}^{-2}$ $\rm (m_a=1~GeV)$ and 
$\frac{\lambda_3}{M_*} < 3.6 \times 10^{-5}~\text{GeV}^{-1}$ $\rm (m_N=1~GeV)$, respectively. They are
somewhat better than the limits that we obtained in this study.  Nevertheless, if we 
could further make use the full missing energy spectrum, we might obtain more competitive
bounds.

Overall, this comparative analysis reveals that, while sensitivity to ALPs and 
sterile neutrinos improves with higher luminosity and systematic control, 
ALPs consistently exhibit better sensitivity at equivalent mass and energy configurations, 
particularly at lower masses and in the absence of systematic uncertainties.

%% file: subtex/Appendix_ALP.tex
\section{ALP Signal Event selection in the Merged region $M_{ET}>500$ GeV, for LHC and HL-LHC}\label{App1}

In this section, we present the cutflow tables for the LHC (Table~\ref{cutflow13_scaled}) and HL-LHC (Table~\ref{cutflow14_scaled}), following the formalism outlined in Section~\ref{sec.4}. The results correspond to the Merged Region with $M_{\text{ET}} > 500~\text{GeV}$. The number of events that survive all selection cuts in the Merged Region at the LHC with $\sqrt{s}=13$ TeV and an integrated luminosity of $\mathcal{L}=139~\mathrm{fb^{-1}}$ is significantly lower compared to the results for the High-Luminosity LHC (HL-LHC) with $\sqrt{s}=14$ TeV and $\mathcal{L}=3000~\mathrm{fb^{-1}}$.

\begin{figure}[h!]
    \centering
    \includegraphics[width=\textwidth]{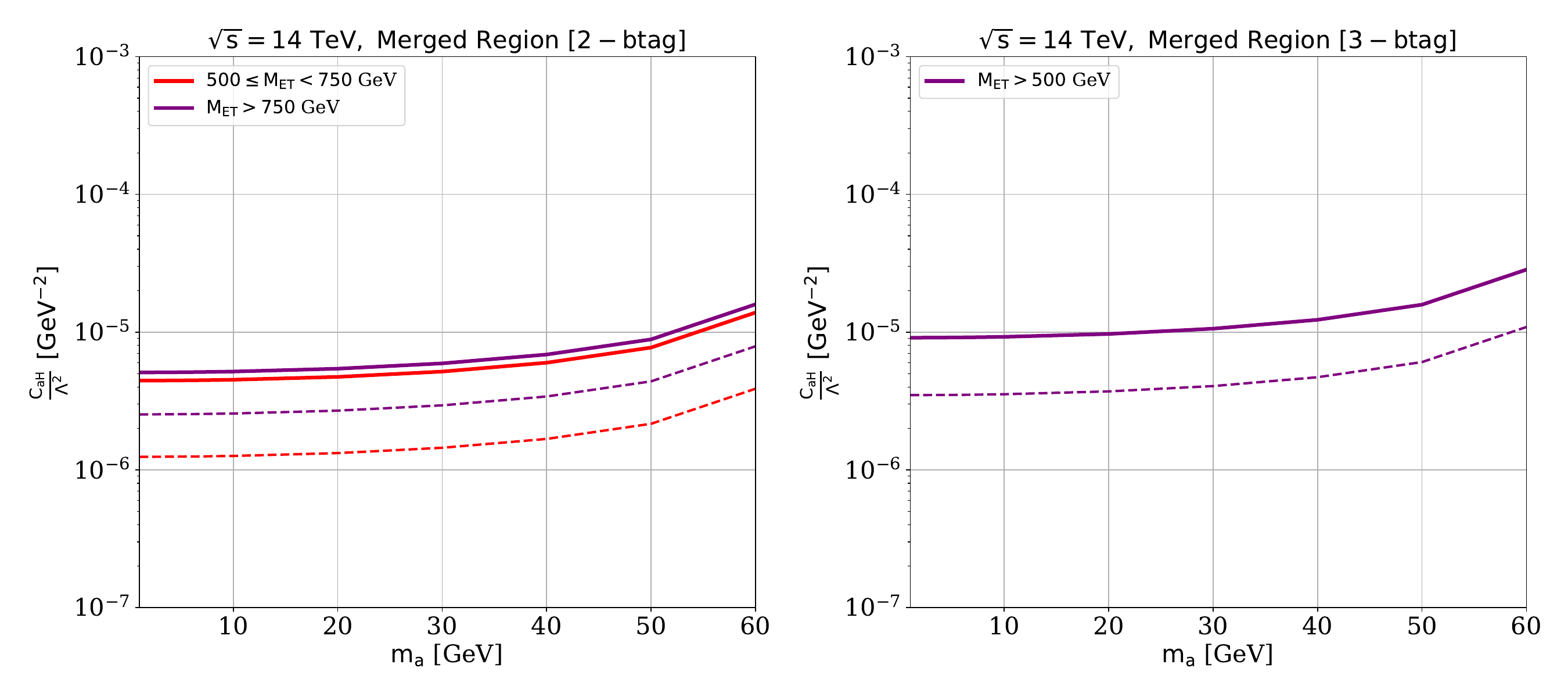}
    \caption{Sensitivity regions (above each curve) at 95\% C.L. for the ALP-Higgs coupling 
    $\frac{C_{aH}}{\Lambda^2}$ at the HL-LHC ($\sqrt{s} = 14$ TeV, $\mathcal{L} = 3000~\mathrm{fb}^{-1}$) in the merged region. 
    \textit{Left Panel}: dashed lines are without systematic uncertainty while the solid lines account for 
    a 20\% systematic uncertainty in the 2 $b$-tag case. \textit{Right Panel}: the same as in the left panel but for 3 $b$-tags.
}
    \label{fig.Sen_Merge}
\end{figure}

Furthermore, we present the exclusion regions at the 95\% C.L. for the ALP-Higgs coupling $\frac{C_{aH}}{\Lambda^2}$ in the Merged Region FIG.~\ref{fig.Sen_Merge}. In the left panel, solid lines represent the exclusion limits for the ALP-Higgs coupling at the LHC with $\sqrt{s}=14$ TeV and $\mathcal{L}=3000~\mathrm{fb^{-1}}$, while the dashed lines incorporate a 20\% systematic uncertainty in the 2 $b$-tag case. Similarly, in the right panel, solid lines depict the exclusion limits for the same center-of-mass energy and integrated luminosity, with the dashed lines reflecting a 10\% systematic uncertainty in the 3 $b$-tag case.

\begin{table}
\centering
\begin{ruledtabular}
\begin{tabular}{lccc}
\textbf{Cut} & \textbf{$m_a = 1$ GeV} & \textbf{$m_a = 30$ GeV} & \textbf{$m_a = 60$ GeV} \\
\hline
Total events                               & 1181.40  & 899.45   & 124.86 \\
$M_{ET} > 150$                             & 538.29   & 406.82   & 56.72  \\
Lepton veto                                & 538.18   & 406.77   & 56.70  \\
Muon veto                                  & 537.82   & 406.41   & 56.65  \\
Tauon veto                                 & 505.24   & 380.34   & 53.02  \\
$\Delta \phi(jet_{123}, M_{ET}) > 20^\circ$ & 463.63   & 348.60   & 48.56  \\
$M_{ET} > 500$                             & 9.14      & 7.45     & 1.00  \\
$\geq 2$ b-tagged variable-R jets             & 3.17      & 2.88     & 0.42  \\
$m_{b\bar{b}}$ mass selection               & 2.55      & 2.28     & 0.33  \\

\hline
\textbf{2-btag} & & & \\
$500 \leq M_{ET}<750$             & 1.98    & 1.79    & 0.27   \\
$M_{ET}>750$                     & 0.37    & 0.35    & 0.037   \\

\hline
\textbf{3-btag} & & & \\
$M_{ET}>500$                     & 0.23    & 0.12    & 0.019   \\

\end{tabular}
\end{ruledtabular}
\caption{ 
Same as Table~\ref{cutflow13}, but in the merged region with $M_{ET}>500$ GeV.}
\label{cutflow13_scaled}
\end{table}

\begin{table}
\centering
\begin{ruledtabular}
\begin{tabular}{lccc}
\textbf{Cut} & \textbf{$m_a = 1$ GeV} & \textbf{$m_a = 30$ GeV} & \textbf{$m_a = 60$ GeV} \\
\hline
Total events                               & 30144.23  & 22950.11  & 3185.93 \\
$M_{ET} > 150$                             & 13953.86  & 10494.80  & 1474.31 \\
Lepton veto                                & 13951.47  & 10493.88  & 1474.05 \\
Muon veto                                  & 13938.56  & 10486.71  & 1473.16 \\
Tauon veto                                 & 13055.50  & 9794.66   & 1377.31 \\
$\Delta \phi(jet_{123}, M_{ET}) > 20^\circ$ & 11941.38  & 8950.26   & 1263.31 \\
$M_{ET} > 500$                             & 245.97    & 184.70    & 26.82   \\
$\geq 2$ b-tagged variable-R jets             & 87.42     & 75.27     & 9.82    \\
$m_{b\bar{b}}$ mass selection              & 68.12      & 61.96      & 7.96    \\

\hline
\textbf{2-btag} & & & \\
$500 \leq M_{ET}<750$                     & 53.65    & 54.16    & 6.18   \\
$M_{ET}>750$                     & 10.85    & 4.13    & 1.01   \\

\hline
\textbf{3-btag} & & & \\
$M_{ET}>500$                     & 3.61    & 3.67    & 0.76   \\

\end{tabular}
\end{ruledtabular}
\caption{Same as Table~\ref{cutflow14}, but in the merged Region with $M_{ET}>500$ GeV.}
\label{cutflow14_scaled}
\end{table}

%% file: subtex/Appendix_St.tex
\section{Sterile Neutrino Signal Event selection in the Merged region $M_{ET}>500$ GeV, for LHC and HL-LHC}\label{App2}

In this section, we provide the cutflow tables for the LHC (Table~\ref{cutflow13_Merged}) and HL-LHC (Table~\ref{cutflow14_Merged}), based on the methodology described in Section~\ref{sec.4}. These results pertain to the Merged Region with $M_{\text{ET}} > 500~\text{GeV}$. At the LHC ($\sqrt{s}=13$ TeV, $\mathcal{L}=139~\mathrm{fb^{-1}}$), the number of events passing all selection criteria in the Merged Region is considerably smaller compared to the High-Luminosity LHC (HL-LHC) scenario ($\sqrt{s}=14$ TeV, $\mathcal{L}=3000~\mathrm{fb^{-1}}$).

\begin{figure}[h!]
    \centering
    \includegraphics[width=\textwidth]{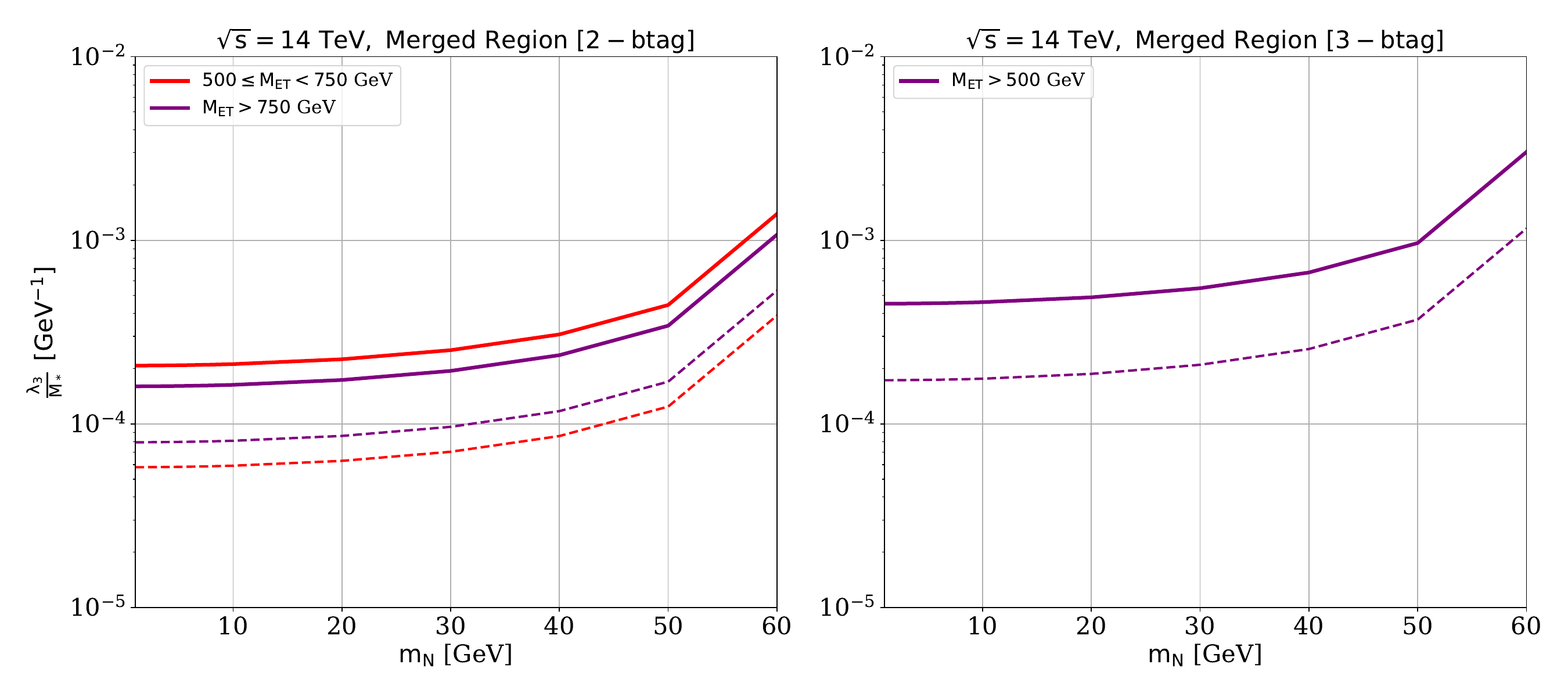}
    \caption{ 
    Same as in FIG.~\ref{fig.Sen_Merge} but for the sterile neutrino-Higgs coupling $\frac{\lambda_3}{M_*}$.
    }
    \label{fig.Sen_Sterile_Merge}
\end{figure}

Additionally, we present the 95\% C.L. exclusion regions for the sterile neutrino-Higgs coupling $\frac{\lambda_3}{M_*}$ in the Merged Region (FIG.~\ref{fig.Sen_Sterile_Merge}). The left panel shows exclusion limits for the sterile neutrino-Higgs coupling at the HL-LHC ($\sqrt{s}=14$ TeV, $\mathcal{L}=3000~\mathrm{fb^{-1}}$), with solid lines representing the nominal case and dashed lines incorporating a 20\% systematic uncertainty for the 2 $b$-tag scenario. The right panel displays similar exclusion limits for the 3 $b$-tag scenario, using the same convention for solid and dashed lines.

\begin{table}[t!]
\centering
\begin{ruledtabular}
\begin{tabular}{lccc}
\textbf{Cut} & \textbf{$m_N = 1$ GeV} & \textbf{$m_N = 30$ GeV} & \textbf{$m_N = 60$ GeV} \\
\hline
Total events                               & 79.177 & 53.731 & 1.764 \\
$M_{ET} > 150$                             & 35.850 & 23.835 & 0.798 \\
Lepton veto                                & 35.845 & 23.833 & 0.798 \\
Muon veto                                  & 35.829 & 23.811 & 0.797 \\
Tauon veto                                 & 33.611 & 22.326 & 0.747 \\
$\Delta \phi(jet_{123}, M_{ET}) > 20^\circ$ & 30.286 & 20.451 & 0.686 \\
$M_{ET} > 500$                             & 0.582  & 0.388  & 0.014 \\
$\geq 2$ b-tagged variable-R jets             & 0.212  & 0.146  & 0.006 \\
$m_{b\bar{b}}$ mass selection               & 0.150  & 0.115  & 0.005 \\

\hline
\textbf{2-btag} & & & \\
$500 \leq M_{ET}<750$             & 0.119 & 0.094 & 0.004 \\
$M_{ET}>750$                     & 0.025 & 0.014 & 0.0005 \\

\hline
\textbf{3-btag} & & & \\
$M_{ET}>500$                     & 0.006 & 0.008 & 0.0003 \\

\end{tabular}
\end{ruledtabular}
\caption{
The same as in Table~\ref{cutflow13_St}, but in the merged region with $M_{ET}>500$ GeV.}
\label{cutflow13_Merged}
\end{table}

\begin{table}[t!]
\centering
\begin{ruledtabular}
\begin{tabular}{lccc}
\textbf{Cut} & \textbf{$m_N = 1$ GeV} & \textbf{$m_N = 30$ GeV} & \textbf{$m_N = 60$ GeV} \\
\hline
Total events                               & 2020.26  & 1370.99  & 45.01 \\
$M_{ET} > 150$                             & 929.88   & 618.80   & 20.43 \\
Lepton veto                                & 929.72   & 618.56   & 20.42 \\
Muon veto                                  & 929.09   & 617.95   & 20.42 \\
Tauon veto                                 & 870.36   & 579.08   & 19.10 \\
$\Delta \phi(jet_{123}, M_{ET}) > 20^\circ$ & 795.95  & 530.30   & 17.46 \\
$M_{ET} > 500$                             & 17.57    & 11.04    & 0.38 \\
$\geq 2$ b-tagged variable-R jets          & 6.83     & 3.84     & 0.14 \\
$m_{b\bar{b}}$ mass selection              & 4.81     & 2.80     & 0.11 \\
\hline
\textbf{2-btag} & & & \\
$500 \leq M_{ET}<750$                      & 3.92     & 2.28     & 0.09 \\
$M_{ET}>750$                               & 0.49     & 0.38     & 0.02 \\
\hline
\textbf{3-btag} & & & \\
$M_{ET}>500$                               & 0.40     & 0.14     & 0.004 \\
\end{tabular}
\end{ruledtabular}
\caption{
The same as in Table ~\ref{cutflow14_St}, but in the resolved region with $M_{ET}>500$ GeV.}
\label{cutflow14_Merged}
\end{table}